\documentclass[aps,pra,amsfonts,amsmath,twocolumn,superscriptaddress]{revtex4-1}
\usepackage{graphicx}
\usepackage{amssymb}
\usepackage{amsmath}
\usepackage{color}
\usepackage{amsbsy}
\newcommand{\br}{\mathbf r}
\newcommand{\bR}{\mathbf R}
 
\newcommand{\bz}{\mathbf z}
\usepackage{color}
\newcommand{\cE}{\ensuremath{{\cal E}}}
 
 \definecolor{BrickRed}{cmyk}{0,0.89,0.94,0.28}%%%PANTONE 1805
\definecolor{MidnightBlue}{cmyk}{0.98,0.13,0,0.43}%%%PANTONE 302
\definecolor{DarkGreen}{rgb}{0,0.7,0.1}

\newcommand{\bfx}{{\bf x}}

\newcommand{\cU}{{\mathcal U}}

\newcommand{\bE}{\mathbf E}
\newcommand{\bB}{\mathbf B}

\newcommand{\bM}{\mathbf M}
\newcommand{\bN}{\mathbf N}

\newcommand{\bX}{\mathbf X}
\newcommand{\curl}{\boldsymbol{\nabla} \times}
\newcommand{\veck}{\mathbf{k}}
\newcommand{\veckpe}{\veck_\perp}

\newcommand{\vecx}{\mathbf{x}}

\newcommand{\be}{\begin{equation}}
\newcommand{\ee}{\end{equation}}

\begin{document}
%\tighten
%\preprint{}
\title{Collective charge fluctuations and Casimir interactions\\ for
  quasi one-dimensional metals}

\author{E. Noruzifar}
\affiliation{Department of Physics and Astronomy,
  University of California, Riverside, California 92521, USA}

\author{T. Emig}
\affiliation{Laboratoire de Physique Th\'eorique et Mod\`eles
 Statistiques, CNRS UMR 8626, Universit\'e Paris-Sud, 91405 Orsay,
 France}

\author{U. Mohideen}
\author{R. Zandi}
\affiliation{Department of Physics and Astronomy,
  University of California, Riverside, California 92521, USA}

\date{\today}
\begin{abstract}  
We investigate the Casimir interaction between two parallel metallic cylinders and
  between a metallic cylinder and plate. The material properties of the metallic
 objects are implemented by the plasma, Drude and perfect metal model dielectric functions. 
We calculate the Casimir interaction numerically at all separation distances and analytically 
at large separations. 
The large-distance asymptotic interaction between one plasma 
cylinder parallel to another plasma cylinder or plate does not depend on the material
  properties, but for a Drude cylinder it depends on the dc conductivity $\sigma$. 
At intermediate separations, 
for plasma cylinders 
the asymptotic interaction depends on the plasma wave length $\lambda_{\rm p}$ while 
for Drude cylinders the 
Casimir interaction can become independent of the material properties. 
We confirm the analytical results by the numerics and 
show that at short separations, the numerical results 
approach the proximity force approximation.
\end{abstract}
\maketitle

\section{Introduction}
\label{intro}
Effective interactions between cylinders are
an important parameter in synthesizing and analyzing nanometric systems. 
This is due to the fact that many important nanostructures such as 
carbon nanotubes, nanowires and even the tobacco mosaic viruses 
have cylinderical shapes. 

From the perspective of the experimental Casimir force studies, nano-cylindrical shapes are
an optimal candidate for precision Casimir force measurements, 
in comparison to
spheres for two reasons: (i) their effective area of interaction is larger
\cite{brown05, decca10}, and (ii) mechanical oscillation modes of quasi-one-dimensional 
structures can be probed with high precision \cite{saz04}.

Under many circumstances, van der Waals or
Casimir forces have the dominant contribution in the effective interactions 
of nanostructures, 
which lead to various interesting phenomena in nanosystesm.
For example 
in nanomechanical devices, Casimir interaction causes stiction \cite{serry98,buk01}, 
and thus a good understanding of these forces 
leads to improvements in the design and efficiency of such nanosystems. 
In another example, 
Casimir interactions between single walled carbon nanotubes (SWCNT)
with different chirality become important in separating
 a polydisperse solution of SWCNT in fractions of equal chirality \cite{podgo}.

The applicability of the Casimir interaction is not limited 
to synthetic cylindrical objects. 
There are numerous examples of long 
macromolecular structures with cylindrical shape in nature such as 
 the tobacco mosaic viruses, 
 microtubules of flagella and A-band lattice of 
myosin filaments in cross strained muscles \cite{gf68, parseg72}, and hence 
 knowledge of the interaction between cylindrical shapes is also important 
for the biological sciences.
It should be noted that in some biological systems composed of cylindrical 
particles which are packed in an array, the separation between the particles 
can be several times larger than the diameter of the cylinder \cite{gf68}.

The Casimir interaction per unit length for two parallel perfectly
conducting cylinders or a plate and cylinder at a separation distance $d$ 
is ${\cal E}/L \sim \hbar c/d^2$, up to a logarithmic factor
\cite{emig06,rahi08}. It decays only slowly compared to the retarded
interaction ${\cal E}/L \sim \hbar c R^4/d^6$ between two insulating cylinders
that do not support {large-scale} collective fluctuations {\cite{rahi09}}.

It has been demonstrated that Casimir interactions strongly depend on
the combined effects of shape and material properties, see, e.g.,
\cite{umar,Emig:2009fk,Graham:2010uq,Maghrebi:2011kx,zandi10,noru11}. The interplay is particularly strong for quasi
one-dimensional conducting materials due to strongly anisotropic
collective charge fluctuations. In addition,
approximations of the Casimir
force between cylinders and plates \cite{decca10} have also shown that the
temperature dependence varies based on the description
of the material properties. Thus there is a need for exact
calculations of the Casimir force for cylindrical shapes taking into
account the realistic material response.

Most studies of
interactions between one-dimensional systems over a wide range of
separations concentrate on perfect conductors and insulators. However, low
dimensionality in combination with finite conductivity and plasmon
excitations should give rise to interesting new effects that might be
probed experimentally using, e.g., the coupling to mechanical
oscillation modes. 
The often employed technique for these effects, the
proximity force approximation (PFA) cannot capture the correlations of
shape and material response since it is based on the interaction
between planar surfaces. 
A number of studies have been performed 
for the short separation regime mainly 
focused on the corrections to the Proximity Force Approximation
\cite{most06, bordag06, lombardo08,holger10}. 

Van
der Waals interaction between cylinders (and plates) {have been
  studied} for certain
frequency dependent permittivities
\cite{barash89,dobson06,dalvit06,emig06,rahi08,drummond07,dobson09}. 
In one of the earliest study, the van der Waals interaction 
has been calculated 
between two parallel thin filaments described by 
one dimensional (1D) 
plasmon and electromagnetic excitations \cite{barash89}. 
This work predicts asymptotic forms of the interaction 
energies at large separations accurately 
but the range of validity of the asymptotics 
{remain unclear.}

In another work {\cite{dobson06}}, the Casimir interaction is obtained 
for conducting cylinders described by 
delocalized coupled 1D plasmons at zero temperature. 
The response function of the plasmons 
are given by the random phase approximation (RPA). 
The specific choice of RPA has the advantage that locality, 
additivity and $R^{-6}$ contributions are not involved in the calculations.
The energy is obtained by using the mode summation method, which is equal to 
the sum of the separation dependent zero-point plasmon modes. 
The Casimir energy is 
attractive and decays as $\sim d^{-2} $ apart from a logarithmic part. 
This result was later confirmed by a quantum Monte Carlo simulation \cite{drummond07}. 
Using the same material description and calculation technique, 
the large-distance Casimir energy 
was obtained for crossed wires with a small crossing angle. 
For conducting and semiconductor wires, 
apart from the logarithmic and angular parts, 
the large-distance interaction energy decays as $\sim d^{-1} $ and $\sim d^{-4}$, respectively \cite{dobson09}.  

In a completely different approach, employed for perfectly conducting 
cylinders and plate, the Casimir energy is calculated for all separations 
using a path integral representation 
for the effective action which yields a trace formula for the
density of states \cite{emig06, rahi08}. 
Furthermore, the Casimir interaction between a SWCNT and a plate is  
studied for large and short separation regimes using the Lifshitz formula \cite{most06}.

The {full} interplay between shape and material effects is not
transparent in the previous studies as they are limited either to perfect
metals or to asymptotic limits.
Here, we employ the scattering approach to 
investigate the Casimir interaction between 
parallel metallic {(circular)} cylinders,
and a metallic cylinder {and a} metallic plate. 
The material properties of the objects are described  either by the plasma or the Drude
dielectric function. 
Some of the results {have been}
reported in Ref.~\cite{noru11}. 

The {outline}  of this work is as follows: in Sec.~\ref{sec:method}, 
we {summarize} the 
scattering method and the {assumed} material properties 
{of the cylinders.} 
In Sec.~\ref{sec:large-separation}, we 
obtain analytical results for the interaction at
distances much larger than {the cylinder radii}. 
In Sec.~\ref{sec:numerics} the Casimir interaction is calculated {numerically}
for different material properties {over a wide range of separations}. 
Section \ref{summary} is the summary.
\section{Method}
\label{sec:method}
We consider the two following systems 
({i}) two infinitely long parallel cylinders, and 
({ii}) an infinitely long cylinder parallel to an infinite plate. Assuming placed in vacuum,
we calculate the Casimir interaction in these two systems employing the scattering formalism \cite{rahi09}. 
The Casimir energy of two objects at zero temperature is given by 
the general expression
\begin{equation}
\label{eq:energy_gen}
\cE = \frac{\hbar c}{2\pi} \int_0^\infty d \kappa\, 
\ln \det ({\bf 1} - \mathbb{N} ) \, ,
\end{equation}
where $\kappa$ is the Wick-rotated frequency and the matrix 
$\mathbb N$ 
factorizes into the scattering amplitudes (T-matrices) 
and translation matrices that describe the coupling between 
the multipoles on distinct objects. While the material properties {and
  shapes} of the 
objects are contained in the T-matrices, the distance between objects is encoded in translation matrices. 

To implement the material properties, we 
consider plasma, Drude and perfect metal cylinders 
with magnetic permeability $\mu= 1$.  The Drude 
dielectric function {on the imaginary frequency axis} is
\begin{equation}
\label{di-fun}
\epsilon (i c \kappa) = 
1+ \frac{(2\pi)^2}{(\lambda_{\rm p}\kappa)^2+\lambda_{\sigma}\kappa/2}\,,
\end{equation}
with {conductivity $\sigma$} and $\lambda_\sigma = 2\pi c/\sigma$.
Equation (\ref{di-fun}) reproduces the plasma model for $\lambda_{\sigma}\to 0$.

Since the matrix $\mathbb N$ differs for parallel cylinders and cylinder-plate systems, 
in the {following}
we {describe} $\mathbb N$ for both setups.
\subsection{{Two} Parallel Cylinders}
\label{subsec:method-cyls}
Consider two infinitely long, parallel cylinders with equal radii
$R$ and {with their axes} separated 
by a distance $d$ and aligned along 
the $z$-axis. The matrix $\mathbb N$ is diagonal in the
  $z$-component $k_z$ of the wave vector due to translational symmetry.
The matrix elements
 for {electric (E) and
  magnetic (M)} polarizations ($\alpha,~\beta = E,~M$) and 
partial waves $m$ and $m'$ are
\begin{equation}
  \label{eq:N_2_cyl}
  {\mathbb N}_{k_zmm'}^{\alpha\beta}=\sum_{\gamma=E,M} T_{k_z m}^{\alpha\gamma} 
\sum_{n=-\infty}^\infty\cU_{k_z mn}^{12} T_{k_z n}^{\gamma\beta} \cU_{k_z nm'}^{21} \,, 
\end{equation}
{with} $T$ 
the cylinder $T$-matrix, see Appendix~\ref{cyltmatrix}. The 
translation matrix $\cU^{12}$ 
{relates regular cylindrical vector waves to outgoing cylindrical
  vector
waves, see Appendix \ref{translation-matrix}.}
{The translation matrices do not couple different polarizations
and for both $E$ and $M$-polarization,  their matrix elements are
given by}
\begin{eqnarray}
  \label{eq:B-matrix-WR}
  {\mathcal U}^{12}_{k_zn n'} = (-1)^{n'}\, K_{n-n'} \left(p\, d\right),\nonumber\\
  {\mathcal U}^{21}_{k_zn n'} = (-1)^{n-n'}\,{\mathcal U}^{12}_{k_zn n'}\;,
\end{eqnarray}
with
 $p=\sqrt{\kappa^2+k_z^2}$ and 
 $K_n(x)$ the modified Bessel function of the second kind.

Since {$\mathbb N$ is} 
diagonal in $k_z$ 
the determinant in Eq.~(\ref{eq:energy_gen}) factorizes 
into determinants at fixed $k_z$, 
{and the sum over $k_Z$ moves in front of the logarithm.}
After taking the continuum limit,
$\sum_{k_z} \to \frac{L}{2\pi} \int_{-\infty}^\infty dk_z$, 
the energy {per unit length $L$ becomes} 
\begin{equation}
  \label{eq:energy_genkz}
  \frac{\cE}{L} = \frac{\hbar c}{4\pi^2} \int_0^\infty d \kappa 
\int_{-\infty}^{\infty} d k_z
\ln \det ({\bf 1} - \mathbb{N} ) \, .
\end{equation}
{Here the determinant is only over the discrete partial wave
  index $n$.}
\subsection{Cylinder {-- Plate} }
\label{subsec:method-cylpl}
Next we consider a cylinder with radius $R$
parallel to a plate. 
We assume that the cylinder is aligned along the $z$ axis 
and the plate is in the $y-z$ plane.
The distance from the center of the cylinder to the 
plate is $d$. The matrix $\mathbb N$ for this geometry is
\begin{equation}
\label{eq:N2_cylpl}
\mathbb{N}_{k_z m m'}^{\alpha \beta} =  \sum_{\gamma=E, M} {{T}}_{k_z m}^{\alpha\gamma} \,{\mathbb{M}_{k_z m m'}^{\gamma\beta}}\,, 
\end{equation}
with
\begin{multline}
\label{Mmatrix}
\mathbb{M}_{k_z m m'}^{\gamma \beta} = 
 \int_{-\infty}^{\infty} dk_y \,
\frac{e^{-2d\sqrt{{\bf k}_{\bot}^2+ \kappa^2}}}{2\sqrt{{\bf k}_{\bot}^2+\kappa^2}} \\
\times \sum_{\gamma'=E,M} D_{ k_z m\gamma,{\bf k}_{\bot}\gamma'}
\;T^{\gamma'}_{{\bf k}_{\bot}} \;
 D_{{\bf k}_{\bot}\gamma', k_z m' \beta}^{\dagger}\;(1-2\delta_{\gamma',\beta}) \,,
\end{multline}
where $k_y$ is the {$y$} component of the wave vector,  
${\bf k_\bot}\equiv (k_y,k_z)$, the matrix ${D}_{n k_z \alpha,{\bf k}_{\bot}\beta}$ 
converts vector plane wave functions and cylindrical vector wave functions, see Appendix~\ref{conv}, and
$T^{\beta}_{{\bf k}_{\bot}}$ is the dielectric plane $T$-matrix presented in Appendix~\ref{platetmatrix}.
The energy of this system
{can also be obtained by Eq.~(\ref{eq:energy_genkz})} as $\mathbb N$ is diagonal in $k_z$,
Note that the determinant is not related to $k_z$; thus, we 
suppress all the $k_z$ indices in what follows.
\section{large-distance asymptotic Casimir energies}
\label{sec:large-separation}

To find the asymptotic form of the Casimir interaction at large separations
$d \gg R$, one needs to obtain the T-matrix expressions 
for a cylinder and a plate. 
Using the dielectric function given in Eq.~(\ref{di-fun}), 
the asymptotic form of the cylinder T-matrix elements {for $E$ polarization} and $n=0$ 
at small frequencies 
($\kappa \ll 1$, $k_z/\kappa$ fixed) reads (see Eq.~(\ref{eq:T-matrix-elements-ee})),
\be
\label{tmatrix_cyl}
T_0^{EE}  \approx -\frac{p^2R^2}{
C(\kappa) - p^2R^2 \ln(p R/2)}\,,
\ee
where $C(\kappa)$ depends on the dielectric properties of the cylinder.  For a perfect metal cylinder $C(\kappa)=0$ and for a plasma
cylinder $C(\kappa) \approx {\lambda_{\rm p}}^2 \kappa^2\,
/(2\pi^2 )$ if the plasmon oscillations cannot build up transverse to
the cylinder axis as the diameter is too small, i.e., $R\ll
\lambda_p$.
In the opposite limit $R \gg \lambda_p$, 
we reproduce the perfect metal form of the T-matrix, i.e. 
$C(\kappa) \approx 0$. 
For the Drude model, $C(\kappa)={\lambda_\sigma \kappa}/{(4\pi^2)}$ if
$\kappa \ll \lambda_\sigma/\lambda_p^2$, $1/\lambda_\sigma$. The first
of the two conditions implies that Drude behavior dominates over
plasma behavior, i.e., the second term in the denominator of
Eq.~(\ref{di-fun}) is larger than the first term. The second
condition ensures that the Drude dielectric function is large compared
to one, i.e., metallic behavior is pronounced.
At small frequencies $\kappa$ but fixed $k_z/\kappa$, for Drude cylinders $T^{EE}_0 \sim \kappa$, while for plasma
and perfect metal cylinders $T^{EE}_0 \sim 1$.  Since
$T^{MM}_0\sim\kappa^2$, $T^{EM}_0=T^{ME}_0=0$ and higher order
elements associated with $n\ne 0$ scale as $\kappa^{2|n|}$, we consider only the
$T_0^{EE}$ elements at large separations.

The Casimir interaction between conducting cylinders is 
intricate and no simple analytical expression that applies to all
 {distances} can be obtained. However, using
Eqs.~(\ref{eq:energy_genkz}), (\ref{eq:N_2_cyl})  and \eqref{eq:N2_cylpl} along with
$T_0^{EE}$ given in Eq.~(\ref{tmatrix_cyl}), the asymptotic
interaction at large separations, $d \gg R$, can be evaluated in
various limiting cases. 

To derive the large-distance asymptotic Casimir potential energy, we employ the identity $\log\det = {\rm Tr}\log$ and
expand the integrand in Eq.~(\ref{eq:energy_genkz}) 
{in powers of $\mathbb N$, corresponding to a multiple scattering
  expansion}. 
 The one-scattering approximation{, sufficient for
  large distances, yields}
\begin{equation}
\label{eq:energy_exp}
\frac{\mathcal E}{L} = -\frac{\hbar c}{4\pi^2}\int_0^{\infty} d\kappa 
\int_{-\infty}^{\infty} dk_z\,  {\rm Tr}[\mathbb N] + ...\,.
\end{equation}
The {element} $\mathbb N_{00}$ of the $\mathbb{N}$-matrix
{yields} the dominant contribution 
{to the} Casimir energy {at large
 distances since higher order elements involve higher powers of
 $\kappa$. To this end, we will consider only this term for the rest of this section.}
 
\subsection{Parallel cylinders}
For two parallel cylinders, considering the fact that at large separations 
$T^{EE}_0$ {yields} the dominant 
contribution, the {trace of the} matrix $\mathbb{N}$ is approximated by 
\begin{eqnarray}
\label{n00-cyls}
{\rm Tr}[\mathbb{N}{^{EE}_{00}}] = T^{EE}_{0}\,\cU_{00}^{12} \,T^{EE}_{0}\, \cU_{00}^{21} \,. 
\end{eqnarray}
Using Eq.~(\ref{n00-cyls}) in Eq.~(\ref{eq:energy_exp}) and 
{changing integration} to polar coordinates 
$\kappa = \rho \cos(\theta)/d$ and $k_z = \rho \sin(\theta)/d$, we obtain
\begin{multline}
\label{eq:energy_exp_cyls}
 {\mathcal E} = -\frac{\hbar c L}{2\pi^2 d^2\ln^2(2d/R)} \\ \times\int_0^{\infty} d\rho 
\int_{0}^{\frac{\pi}{2}} d\theta\,\frac{\rho
  K_0^2(\rho)}{(1+C_1(\rho,\theta))(1+C_2(\rho,\theta))}\, ,
\end{multline}
{where $C_i(\rho,\theta)$ describes the material properties of the
cylinder $i$.}  For a perfect metal cylinder $C_i(\rho,\theta)=0$, 
for a plasma cylinder $C_i(\rho, \theta) {=} \xi \cos^2(\theta)$ with 
\begin{equation}
\label{eq:ksi}
\xi = \frac{\lambda_{\rm p}^2}{2\pi^2 R^2 \ln(2d/R)}\,,
\end{equation}
{in the limit $R \ll \lambda_p$, and} 
for a Drude cylinder $C_i(\rho,\theta) = \xi' \cos(\theta)/\rho$ with 
\begin{equation}
\label{eq:ksip}
\xi'= \frac{\lambda_{\sigma} d}{4\pi^2 R^2\ln(2d/R)}.
\end{equation}
For two {\it perfect metal cylinders} the integral in Eq.~(\ref{eq:energy_exp_cyls}) can easily be 
calculated {and yields}
\begin{equation}
\label{cyls-pf}
\frac{{\mathcal{E}}}{\hbar c L} \approx -\frac{1}{8\pi d^2
  \ln(2d/R)^2}\, ,
\end{equation}
which agrees with the results in Refs.~\cite{emig06,rahi08}.

\begin{figure}
\begin{center}
\includegraphics[width=0.45\textwidth]{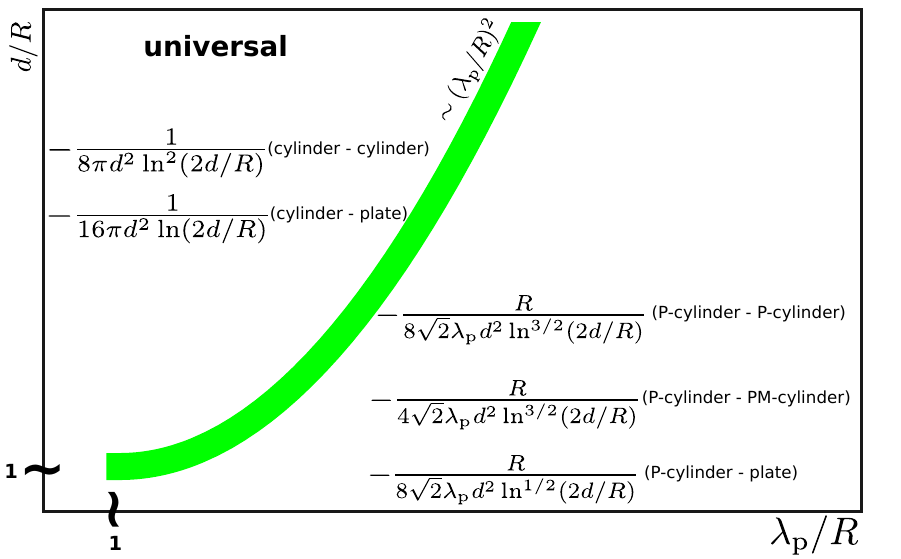}
\includegraphics[width=0.45\textwidth]{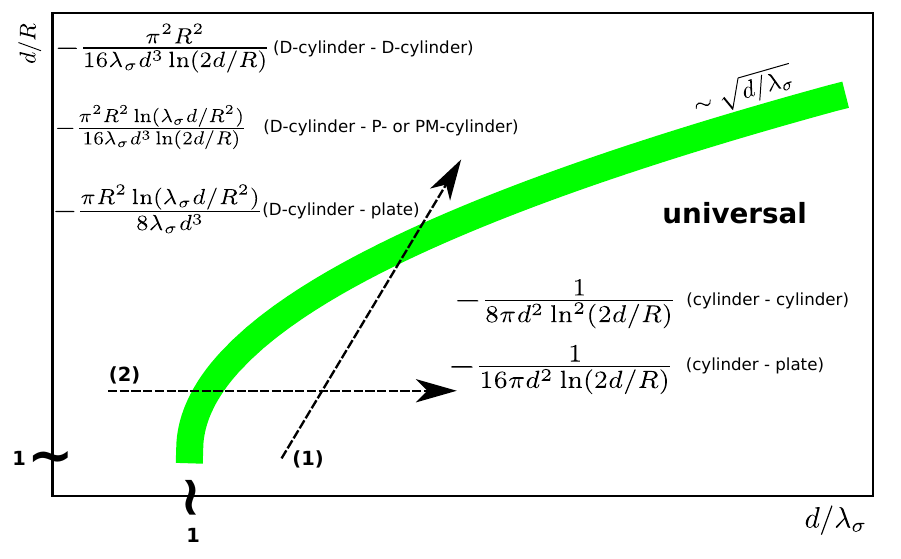}
\caption{\label{crossovers} Summary of the different forms of
  interaction between two cylinders and a cylinder and a plate. Shown are the
  rescaled interaction energies per cylinder length, ${\cal E}/(\hbar c L)$. (a)
  Interaction involving a plasma (P) cylinder with another
  plasma cylinder, a perfect metal (PM) cylinder or a plate. The asymptotic
  results apply sufficiently far away from the separating
  curve $\ln(d/R)\sim (\lambda_p/R)^2$ and for $d/R$, $\lambda_p/R \gg
  1$. (b) Interaction involving a Drude (D) cylinder with
  another Drude cylinder, a plasma cylinder, a perfect metal cylinder or a plate.
  The separating curve is given, up to logarithmic corrections, by
  $d/R \sim \sqrt{d/\lambda_\sigma}$. The shown expressions hold for
  $d/R$, $d/\lambda_\sigma \gg 1$ and $d \gg
  \lambda_p^2/\lambda_\sigma$. Depending on the relative size of length scales,
different regimes can be reached: Arrow (1)
  corresponds to an increasing distance $d$ which ultimately leads to
  a {\it non-universal} interaction. Arrow (2) indicates an
  overall increase of the geometry (i.e., $d/R$ fixed) with constant
  conductivity leading to a {\it universal} interaction.
}
\end{center}
\end{figure}

For {\it plasma cylinders} with plasma wave length $\lambda_{\rm p}$, 
{the}  integrations in 
Eq.~(\ref{eq:energy_exp_cyls}) yield
\begin{equation}
\label{eq:plasmaasym}
 \frac{{\mathcal{E}}}{L} \approx -\frac{\hbar c}{16 \pi d^2
   \ln(2d/R)^2}  f({\xi})\,,
\end{equation}
with
\begin{equation}
\label{eq:fofx}
 f(x) = \frac{x+2}{(x+1)^{3/2}}\,.
\end{equation}
It is important to {consider} 
Eq.~(\ref{eq:plasmaasym}) in two limiting cases for $\xi$, Eq.~(\ref{eq:ksi}). 
In the limit $\xi \ll 1$ or $\ln(2d/R)\gg \lambda_{\rm p}^2/R^2$, 
{the energy simplifies} to the perfect metal energy given in Eq.~(\ref{cyls-pf}), i.e., the interaction between conducting cylinders
is universal {at large distances} in this regime.
However, in the opposite limit $\xi \gg 1$ or equivalently $\ln(2d/R)\ll \lambda_{\rm p}^2/R^2$, 
the Casimir energy {becomes}
\begin{equation}
\label{eq:cyls-pl}
 \frac{{\mathcal E}}{L} \approx -\frac{\hbar c \, R}{8\sqrt{2}\, 
 \lambda_{\rm p}\, d^2 \ln^{3/2}(2d/R)}\, .
\end{equation}
{This shows}
 that the universal form 
is applicable only beyond 
an exponentially large crossover length $d\sim R \exp(\lambda_{\rm
  p}^2/R^2) {\gg\!\!\gg R}$.  Below this scale, {and infact
  in any practical situations,} the interaction is material 
dependent, see Fig.~(\ref{crossovers})a.

For a {\it plasma cylinder} with plasma wave length $\lambda_{\rm p}$ 
parallel to a {\it perfect metal} cylinder {the} integrations in 
Eq.~(\ref{eq:energy_exp_cyls}) yield
\begin{equation}
\label{eq:plasmaasymx}
 \frac{{\mathcal{E}}}{L} \approx -\frac{\hbar c}{8 \pi d^2
   \ln(2d/R)^2} (1+\xi)^{-{{1}\over{2}}}\,.
\end{equation}
Similar to parallel plasma cylinders, we consider two limiting cases for $\xi$. 
In the limit $\xi \ll 1$ or $\ln(2d/R)\gg \lambda_{\rm p}^2/R^2$, 
we obtain the perfect metal energy given in Eq.~(\ref{cyls-pf}), 
and the conducting cylinders'
interaction is universal {at large distances}.
In the opposite limit $\xi \gg 1$ or equivalently $\ln(2d/R)\ll \lambda_{\rm p}^2/R^2$, 
the Casimir energy {becomes}
\begin{equation}
 \frac{{\mathcal E}}{L} \approx -\frac{\hbar c \, R}{4\sqrt{2}\, 
 \lambda_{\rm p}\, d^2 \ln^{3/2}(2d/R)}\, .
\end{equation}

For {\it Drude cylinders} with the characteristic length $\lambda_\sigma$, 
at large separations $d\gg R, \, \lambda_\sigma$, 
we find a
rather distinct behavior that deviates from naive expectations for
universality. {In this case the integrations in
Eq.~(\ref{eq:energy_exp_cyls}) } cannot be performed analytically. 
Therefore we first calculate the {angular} integral
 which 
gives a complicated radial function, and then we expand the resulting radial integral 
{for small and large $\xi'$, Eq.~(\ref{eq:ksip}).}
 The radial integrals 
can be calculated easily in these two limits. 
{For} $\xi' \ll 1$ or {$d \ll R^2/\lambda_{\sigma}$}, 
we reproduce the universal (perfect metal) asymptotic energy of Eq.~(\ref{cyls-pf}).
In the opposite limit $\xi'\gg 1$ or {$d \gg R^2/\lambda_{\sigma}$}, the asymptotic energy reads
\begin{equation}
\label{eq:cyls-dr}
 \frac{\mathcal{E}}{L} \approx - \frac{\pi^2\,\hbar c\, R^2}{16\,\lambda_{\sigma}\, d^3 \ln(2d/R)}\,.
\end{equation}
Similarly, for a {\it Drude cylinder} with the characteristic length $\lambda_\sigma$
 parallel to a {\it plasma (or perfect metal) cylinder}, in the limit of 
$d \gg \lambda_{\sigma} $ and $d \gg \lambda_{\rm p}^2/\lambda_{\sigma}$ 
the asymptotic Casimir energy is
\begin{equation}
 \frac{\mathcal E}{L} \approx -\frac{\pi^2 \, \hbar c \,R^2}{16\,\lambda_{\sigma}\, d^3 \ln(2d/R)}
\ln(\lambda_{\sigma}\,d/R^2)\,.
\end{equation}
These two limiting cases for $\xi'$ are related to 
two different scaling regimes that are separated, up to
logarithmic corrections, by the curve $d/R \sim
\sqrt{d/\lambda_\sigma}$, see Fig.~\ref{crossovers}(b). The
 {unconventional} 
feature corresponds to the fact that the interaction is universal 
{at shorter distances} where $d\ll
R^2/\lambda_\sigma$. If the distance is increased beyond this
crossover scale (with all other length scales kept fixed, see arrow
(1) in Fig.~\ref{crossovers}(b), the interaction becomes material
dependent and, up to logarithmic corrections, scales as
$ R^2/(\lambda_\sigma d^3)$ for a Drude cylinder interacting with another
Drude or a plasma or a perfect metal cylinder.  However, if the radii of the cylinders are increased in the same way as their
distance ($d/R$ fixed, see arrow (2) in Fig.~\ref{crossovers}(b)),
finite conductivity becomes unimportant at large distances and the
interaction assumes the universal form.  An intuitive explanation
of this non-universal large distance behavior is given below. 
{It is important to note that all forms of these metallic
  interactions decay much slower
than the Casimir energy of two} insulating cylinders {which} for $d\gg R$
scales as $\hbar c L R^4/d^6$ with a material dependent coefficient.

\subsection{Cylinder parallel to a plate}
In this section we consider a cylinder with radius $R$ parallel to a plate. 
We show, similar to parallel cylinders,
{the existence of} two different
scaling regimes that are separated by curves given by the
same expressions that we found for two cylinders, see
Fig.~\ref{crossovers}. 
In order to find 
the {asymptotic large distance} interations,
 we employ {again} Eq.~(\ref{eq:energy_exp}).
The trace of the matrix $\mathbb{N}$ in Eq.~(\ref{eq:N2_cylpl}) in the limit of 
large separation $d\gg R$ is approximated by
\begin{eqnarray}
\label{n00-cylpl}
{\rm Tr}[\mathbb{N}{^{EE}_{00}}] = {T}^{EE}_{0}\mathbb{M}^{EE}_{00}\,.
\end{eqnarray}
Note that for perfect metal plates $T^{E}_{{\bf
    k}_{\bot}}=T^{M}_{{\bf k}_{\bot}} = 1$ and for small $\kappa$ at
fixed $k_\perp/\kappa$ one has for the plasma model $T^{E}_{{\bf
    k}_{\bot}}=T^{M}_{{\bf k}_{\bot}} = 1+{\cal O}(\lambda_p \kappa)$
and for the Drude model $T^{E}_{{\bf k}_{\bot}}=T^{M}_{{\bf k}_{\bot}}
= 1+{\cal O}(\lambda_\sigma \kappa)$. {Therefore, at large distances the material desription of the
  plate is unimportant and to leading order in $R/d$ one gets}
\begin{eqnarray}
\label{eq:mee00}
{\mathbb{M}}^{EE}_{00} \approx \int_{-\infty}^{\infty} dk_y\;
\frac{e^{-2d\sqrt{\kappa^2+{\bf k}_{\bot}^2}}}{2\sqrt{\kappa^2+{\bf k}_{\bot}^2}} 
{=} K_0(2pd)\,.
\end{eqnarray}
{Using} Eq.~(\ref{n00-cylpl}) in Eq.~(\ref{eq:energy_exp}),  
and {changing again} considering $\kappa = \rho \cos(\theta)/d$ and $k_z = \rho \sin(\theta)/d$, we obtain
\begin{equation}
 \label{eq:energy_exp_cylpl}
\frac{\mathcal E}{L} = -\frac{\hbar c}{2\pi^2 d^2\ln(2d/R)}\int_0^{\infty} d\rho 
\int_{0}^{\pi\over2} d\theta\,\frac{\rho\, K_0(2\rho)}{1+C(\rho,\theta)}\,,
\end{equation}
{where the functions $C(\rho,\theta)\equiv C_i(\rho,\theta)$ are given by the expressions
  below Eq.~\eqref{eq:energy_exp_cyls}}

For a {\it perfect metal cylinder}, the integrals can be calculated 
in a straight forward manner, resulting in the 
universal energy
\begin{equation}
\label{eq:cylpl-pf}
 \frac{\mathcal{E}}{ L} = -\frac{\hbar c}{16\,\pi\, d^2 \ln(2d/R)}\,.
\end{equation}

For a {\it plasma cylinder} with the plasma wavelength $\lambda_{\rm p}$, 
after performing the radial and angular integrals in 
Eq.~(\ref{eq:energy_exp_cylpl}), we obtain 
\begin{equation}
 \label{eq:plasma_asym_cylpl}
 \frac{\mathcal E}{ L}=-\frac{\hbar c}{16\,\pi\, d^2 \ln(2d/R)}\; g({\xi})\,,
\end{equation}
with $g(x) = (1+x)^{-1/2}$. 
{As in our analysis for parallel cylinders}
{we consider} two different cases for $\xi$ given in Eq.~(\ref{eq:ksi}).
If  $\xi \ll 1$ or similarly $\ln(2d/R)\gg \lambda_{\rm p}^2/R^2$,  we reproduce 
the universal energy, Eq.~(\ref{eq:cylpl-pf}).
In the opposite limit {of exponentially large distances}
$\xi \gg 1$ or {$d \gg
  R \exp(\lambda_{\rm p}^2/R^2)$} 
the Casimir energy 
{ is non-universal} and {we obtain}
\begin{equation}
\color{black} 
\label{eq:cylpl-pl}
\frac{\mathcal E}{ L}=-\frac{\hbar c  \, R}{8\sqrt{2}\,\lambda_{\rm p}\,d^2 \ln^{1/2}(2d/R)}\,.
\end{equation}
For a {\it Drude cylinder} with the characteristic length $\lambda_\sigma$
 parallel to a metallic plate, 
in the limit {$\xi' \ll 1$ or $d \ll R^2/\lambda_\sigma$}, 
{the integrand in Eq.~\eqref{eq:energy_exp_cylpl} becomes
  independent of $\theta$ and}
we reproduce the {\it universal} Casimir energy in Eq.~(\ref{eq:cylpl-pf}).
{Hence, similar to the case of two cylinders, the interaction approaches
  a universal form {\it below} a geometry and material dependent
  crossover distance. This counterintuitive result shall be discussed below.}
In the opposite limit $\xi'\gg 1$ or $d\gg R^2/\lambda_\sigma$,  the 
 asymptotic energy {becomes
  {\it non-universal} and reads}
\begin{equation}
\label{eq:cylpl-dr}
\frac{\mathcal E}{ L}=-\frac{\pi\, \hbar c \, R^2
  \ln(\lambda_{\sigma}\,d/R^2)}{8\,\lambda_{\sigma}\, d^3}\, .
\end{equation}
{Based on the studies described above,  we conclude that the Casimir interaction between a metallic cylinder and a plate decays
slower than that between an}
insulating cylinder and a plane for which the energy scales as 
${\cal E}\sim \hbar c LR^2/d^4$ {for $d\gg R$}
\cite{rahi09}.

\section{Numerical Results}
\label{sec:numerics}

In this section, we compute the Casimir energy based on
Eq.~\eqref{eq:energy_genkz} at zero temperature.  Our results are
obtained by numerical computation of the determinant and the 
integrals over $\kappa$ and $k_z$. Note that for a cylinder
parallel to a plate in addition to the $\kappa$ and $k_z$ integrations,
one has to compute the integral over $k_y$ for each element of the
matrix $ \mathbb N$, see Eqs.~\eqref{eq:N2_cylpl} and \eqref{Mmatrix}.
The matrix $\mathbb N$ (and {hence} the sum over $n$ in Eq.~\eqref{eq:N_2_cyl})
are truncated at a finite partial wave number $n_{\rm max}$.

We chose $n_{\rm max}$  such that the
result for the energy changes by less than 
{$0.01\%$} upon increasing $n_{\rm max}$ by $10$.
The required value
of $n_{\rm max}$ {diverges when} the surface-to-surface 
separation {$h$} between the objects (for cylinders $h=d-2R$ and
for {a} cylinder {and a} plate $h=d-R$) {tends to zero}.  
For
example, for $0.6<h/R<1.0$, we used $n_{\rm max}=21$, whereas for $h/R
=0.6$ and $0.5$, one needs $n_{\rm max}=31$. For $h/R=0.05$ we set the
value $n_{\rm max}=191$. 
\footnote{In the cylinder-plate case, we restricted the numerics to
  $h/R>0.3$ with $n_{\rm max}=41$ due to increasing numerical
  uncertainties in the $k_y$ integration in Eq.~(\ref{Mmatrix})}.

To reach sufficient numerical accuracy in the computation of
$\det({\bf 1}-{\mathbb N})$ we have computed the Bessel functions
with quadruple precision and employed similarity transformations for 
$\mathbb N$ by using the DEGBAL routine of the {\it LAPACK} library 
with quadruple precision \cite{laug}.

\begin{figure}
\includegraphics[width=0.45\textwidth]{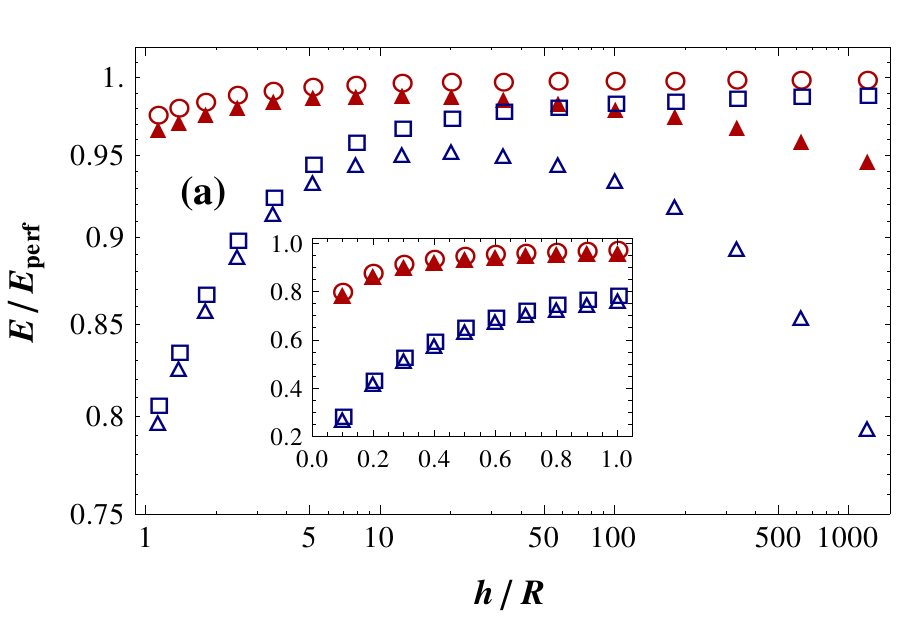}
\includegraphics[width=0.45\textwidth]{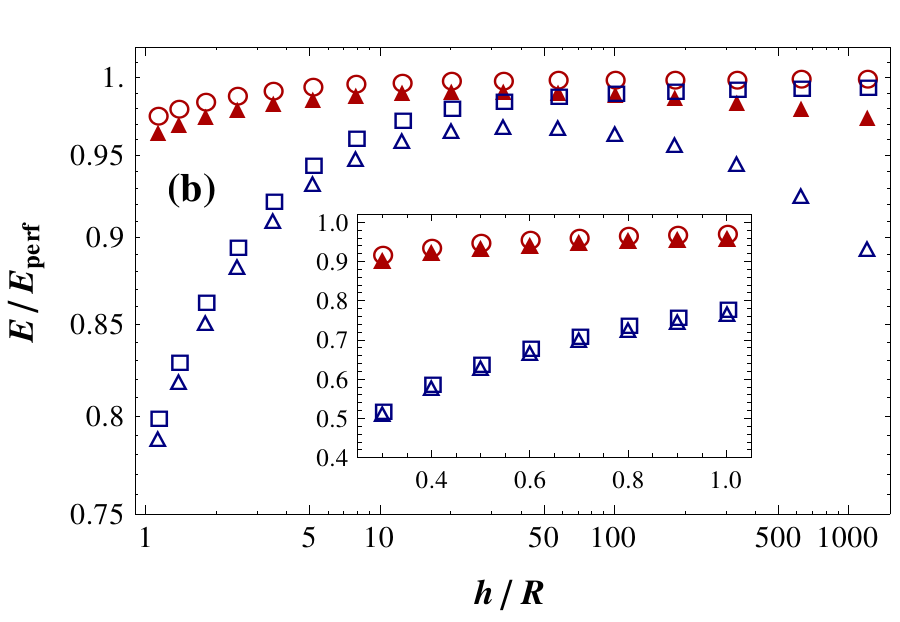}
  \caption{ {Ratio of numerically computed
      energies $E$ for realistic metals and $E_{\rm perf}$ for perfect
      metal} against $h/R$. {(a) two identical cylinders, (b) a cylinder and a
      plate. The parameters} for the plasma model 
  {are} $\lambda_p/R=0.05$, $0.5$ (open circles and open squares
  respetively). {For} the Drude model the same values of
  $\lambda_p/R$ are used; $\lambda_p/R=0.05$, $0.5$(filled and open triangles, respectively) and
  $\lambda_p/\lambda_\sigma= 27.4$. Insets: Short distance range.}
  \label{fig:numerics}
\end{figure}

In Fig.~\ref{fig:numerics}, we show 
{our} numerical results for two parallel cylinders
and also for a cylinder parallel to {a} plate. 
The graphs show the Casimir energies for the Drude
and plasma cylinders, normalized to the energies of perfect 
metal cylinders. For the
numerics we used $\lambda_{\rm p}/R=0.05$ and $0.5$ with $\lambda_{\rm p}/\lambda_{\sigma}=27.4$, corresponding to gold for which $\lambda_{\rm p}= 137$ nm
and $\lambda_\sigma \approx 5$ nm. 
Figure \ref{fig:numerics} clearly {shows} the
material dependence of the Casimir energies.  At large separations,
for the plasma model, the ratios of $E/E_{perf}$ approach one. 
This is due to the fact that $\lambda_{\rm p}/R < 1$ and 
we are in the universal regime, see Fig.~\ref{crossovers}. 
For Drude cylinders, the quantity $\xi'$ determines the behavior of
the curves.  
At $\lambda_{\rm p}/R=0.5$,   one has $\ 5\times 10^{-4} <\xi' <0.06$ and 
for $\lambda_{\rm p}/R=0.05$, $5 \times 10^{-5} <\xi'< 0.006$ for the
range of $h$ shown in Fig.~(\ref{fig:numerics}). 
Since $\xi' \ll 1$, we expect from our asymptotic computations that the Casimir
energy is close to the energy for perfect metal cylinders. Fig.~\ref{fig:numerics}
indeed shows a plateau at intermediate distances that is approaching
the perfect metal energy $E_{perf}$. This approach is better for $\lambda_{\rm p}/R=0.05$
which corresponds to a smaller $\xi'$.  At small distances, none of our
asymptotic results applies and the actual energy is more strongly
redruced compared to $E_{perf}$. With increasing distance, we expect
at $\xi' \sim 1$ a crossover to the non-universal asymptotic energy of
Eq.~\eqref{eq:cyls-dr}. While this crossover is not fully shown in
Fig.~\ref{fig:numerics}, the descrease of the energy ratio
$E/E_{perf}$  with increasing
separation is a precursor of this crossover. The same arguments apply
to  the interaction of a Drude cylinder with a plate. 

We now compare our numerics with the PFA results at short separations
and with the asymptotic results at large separations for both the
plasma and the Drude models.

\subsection{PFA versus numerics}

\begin{figure}
\includegraphics[width=0.4\textwidth]{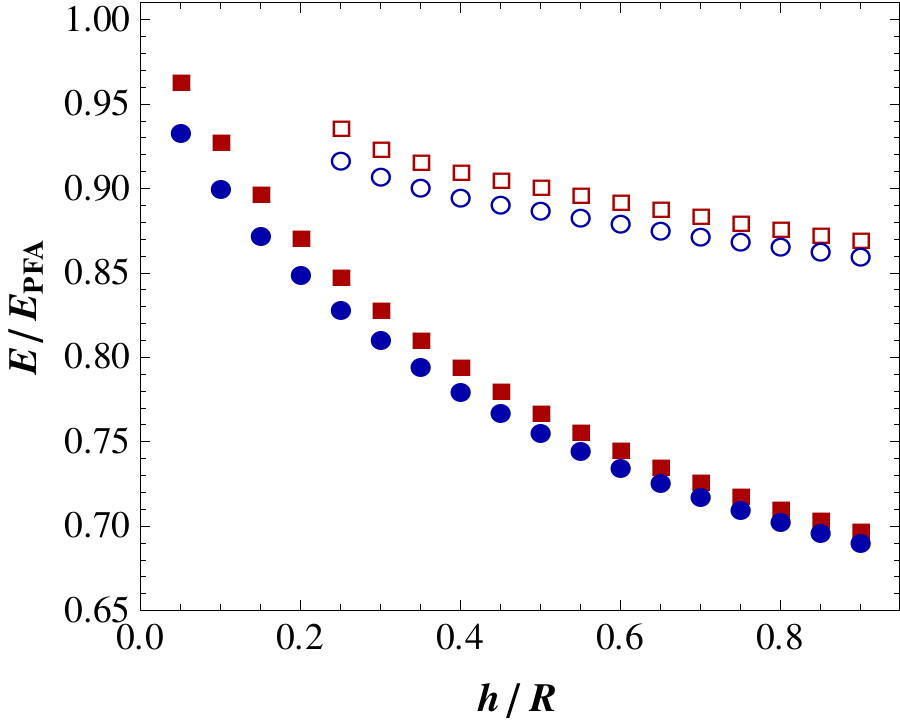}
  \caption{Ratio of the numerical results for the Casimir energy shown
  in Fig.{~\ref{fig:numerics}} and the PFA energy based on the Lifshitz
  theory for the plasma model with $\lambda_p/R=0.05$ (squares) and
  $\lambda_p/R=0.5$ (circles). The ratio is shown as a function of the
  surface-to-surface distance $h$. Filled and empty shapes representing the data 
for parallel cylinders and cylinder--plate, respectively.   }
  \label{fig:pfa}
\end{figure}

The PFA
energy is obtained by integrating the PFA force $F=2\pi R E_{\rm
  plates}(h)$ with respect to $h$, where $E_{\rm plates}(h)$ is
the energy of two parallel plates at distance $h$ given by the
Lifshitz formula \cite{Lifshitz56} using the dielectric
function {of Eq.~\eqref{di-fun}}.  Fig.~\ref{fig:pfa} 
shows the {numerically computed} Casimir energy 
{for $\lambda_{\rm p}/R=0.5$ and
  $0.05$},
normalized {to} the PFA energy for parallel cylinders and
{a} cylinder parallel to a plate. 
We find similar results for the Drude model. 
The energies associated with the Drude model are not shown here since they collapse 
on the data for the plasma model at short separations. 
Our data support the consistency of the PFA in the limit of vanishing separations. 

\subsection{Asymptotics versus numerics}
\label{subsec:asymvsnum}

\begin{figure}
\includegraphics[width=0.4\textwidth]{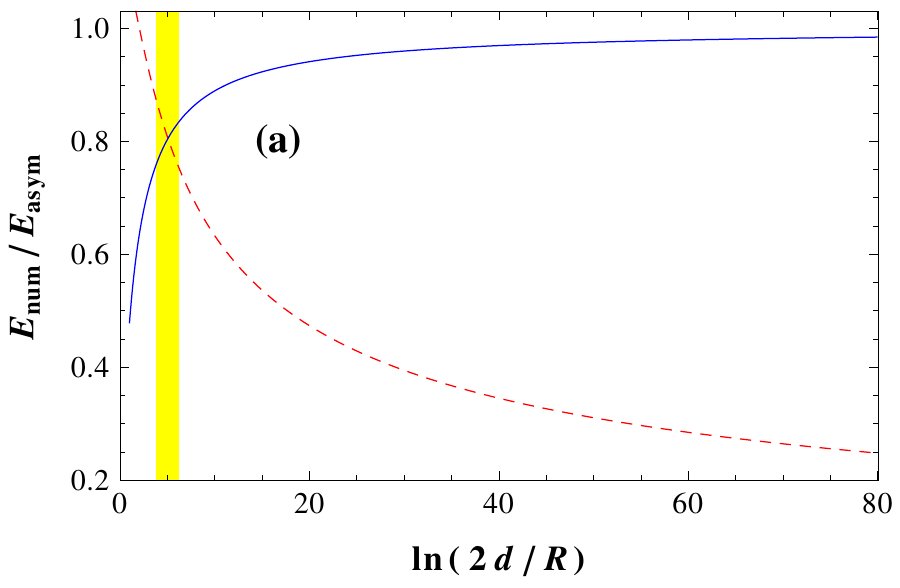}
\includegraphics[width=0.4\textwidth]{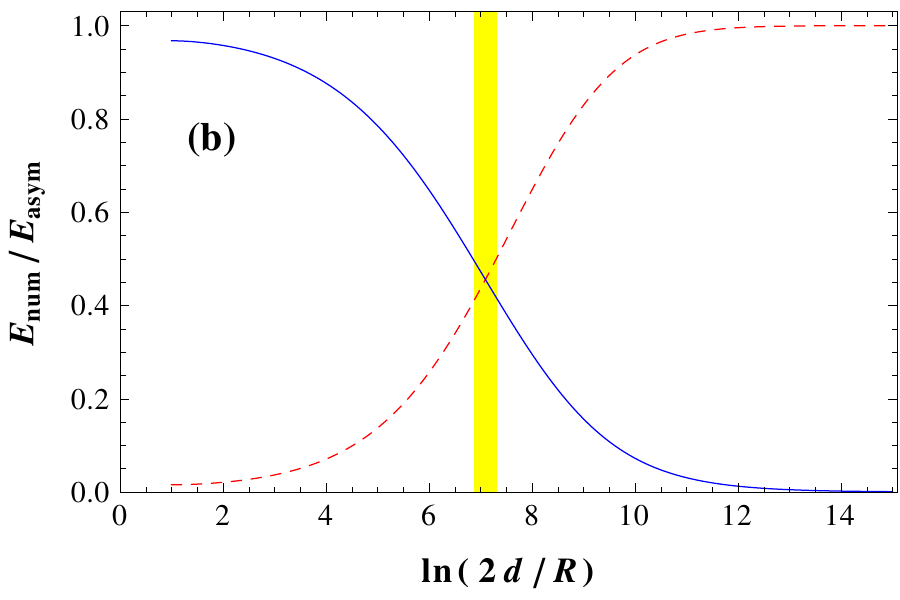}
   \caption{ \label{fig:cylsasym} {Two identical cylinders:} Ratio
     of the numerically computed energy {for the plasma model (a)
       and the Drude model (b)} and the {corresponding} 
  universal (blue solid line) and non-universal   
  (red dashed line) asymptotic results 
  with $\lambda_{\rm
    p}/R=5$ and $\lambda_{\sigma}=\lambda_{\rm p}/27.4$. 
The universal asymptotic energy is given by
Eq.~(\ref{cyls-pf}) and non-universal ones are given by Eqs.~\eqref{eq:cyls-pl} and \eqref{eq:cyls-dr}. 
The yellow strip shows the crossover region between the Drude and plasma asymptotic energies. 
}
 \end{figure}
\begin{figure}
\includegraphics[width=0.4\textwidth]{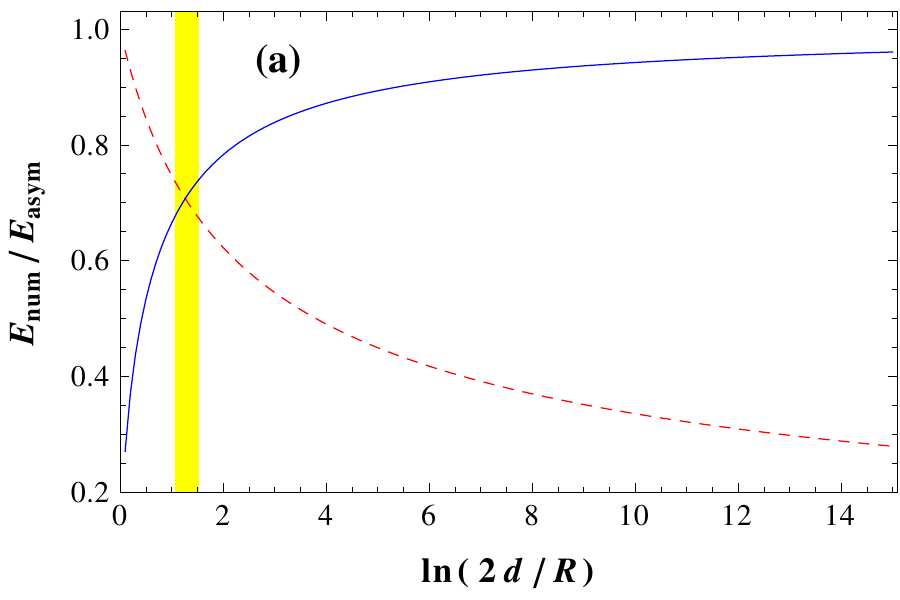}
\includegraphics[width=0.4\textwidth]{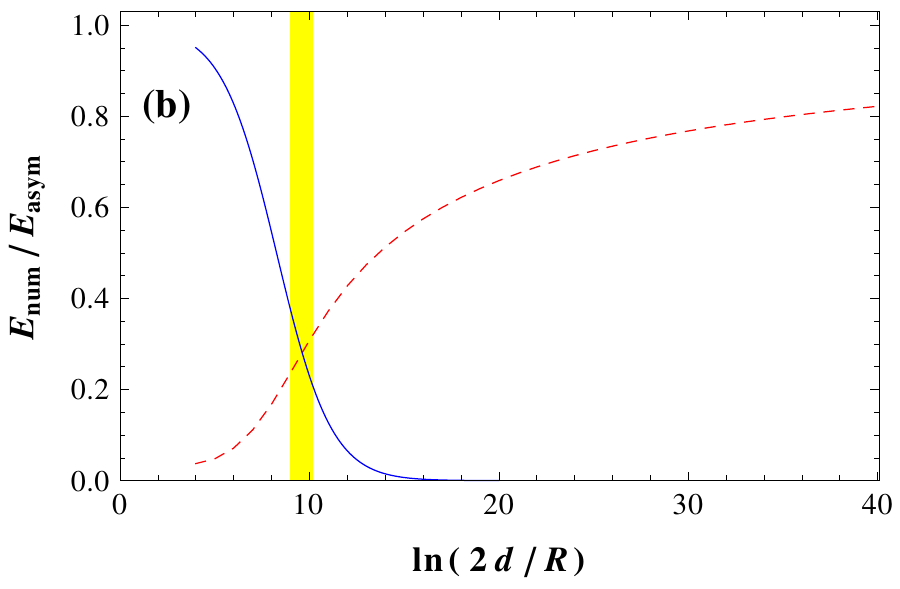}
    \caption{ \label{fig:cylplasym} {Same as
Fig.~\ref{fig:cylsasym} but for a cylinder and a plate with the universal asymptotic energy given by
Eq.~(\ref{eq:cylpl-pf}) and non-universal ones by Eqs.~\eqref{eq:cylpl-pl} and \eqref{eq:cylpl-dr}.}}
  \end{figure}

Figures \ref{fig:cylsasym} and \ref{fig:cylplasym} 
show the ratio of the computed energies 
and the corresponding asymptotic results
(universal and non-universal regimes)
versus $\ln(2d/R)$ for two cylinders and {a} cylinder parallel to
{a} plate, respectively. 
The parameter {that couples shape (radius) and material
properties} is chosen as
$\lambda_{\rm p}/R=5$. 
Figures \ref{fig:cylsasym}(a) and  \ref{fig:cylplasym}(a)
{show that}
for plasma cylinders, at intermediate separations, the energy 
normalized to the non-universal {asymptotic energy} 
is approaching unity whereas at
asymptotically large separations the energy normalized to the
universal {asymptotic energy} is tending to unity. 
On the other hand, 
Figs.~\ref{fig:cylsasym}(b) and  \ref{fig:cylplasym}(b)
{show that}
for Drude cylinders, at intermediate separations, the energy 
normalized to the universal {energy} is approaching unity whereas at
asymptotically large separations the energy normalized to the
non-universal {energy} is tending to unity.  
These figures confirm the validity of
the crossover regime shown in Fig.~\ref{crossovers}.

\section{Summary}
\label{summary}
In summary, we have calculated the Casimir force between two metallic
cylinders and a metallic cylinder parallel to a plate. The energy 
is calculated numerically for {a large range of} separations. We also find asymptotic 
energies for large separations and confirmed their validity with the 
numerical results. Furthermore, we showed that the numerics 
{tend to the} PFA 
energies at short separations.

The interesting phenomenon in our results is that the Casimir
interaction involving Drude cylinders approaches {a}
universal {form of} interaction at intermediate separations and becomes {\it
  non-universal} (material dependent) at {larger} distances $d
\gtrsim R^2/\lambda_\sigma$.  This behavior can be explained in terms
of the size of the collective charge fluctuations in a Drude metal.
However if $\lambda_{\rm p} \ll R$ then plasma oscillations are supported by the cylinders
and at asymptotically large separations the interaction energy does show universality. 

Based on recent experiments, the interactions between
metals might not be consistent with the Drude model \cite{Decca07}. 
The asymptotic energies that we found in this work can be used to provide a 
clearer distinction between the Drude and plasma model predictions 
as compared to two plates or a plate and
sphere \cite{zandi10}. 
An estimate of the interaction between 
two gold cylinders with $R=10$nm, length
$L=100\mu$m, $\lambda_p=137$nm and $\lambda_\sigma=5$nm at a distance
$d=200$nm yields a force of $\approx 1$pN within the plasma description
and $\approx 27$pN within the Drude model. These forces are
in the experimentally detectable regime.

The significant feature of the interaction between a Drude cylinder with
another Drude cylinder or a plate is that upon increasing the
separation, the interaction can move from a universal regime to a 
non-universal one.  This behavior can be understood from the wave
equation for the electric field inside a Drude cylinder.  For
imaginary frequencies $\omega=ic\kappa$, the Helmholtz operator
$\nabla^2 + \epsilon(\omega) (\omega/c)^2$ for a good Drude conductor
becomes $\nabla^2 - 8\pi^2 \kappa/\lambda_{\sigma}$. We are interested
in the maximal wave length of the field and hence charge fluctuations for a
given $\kappa$. With the smallest transverse wave vector $k_x$,
$k_y\sim 2\pi/R$ we find 
the dispersion relation
\begin{equation}
  \label{eq:dispersion}
  |k_z| \sim R^{-1}\sqrt{\kappa/\kappa_c -1}\, , \quad \kappa_c =
  \lambda_\sigma/R^2 \, .
\end{equation}
Hence, collective charge fluctuations on arbitrarily large scales
exist only for $\kappa > \kappa_c$ which is a consequence of
dimensionality that does not appear in the absence of transverse
constraints ($R\to\infty$).  For $\kappa<\kappa_c$ charge fluctuations
break up into clusters of typical size $\sim
R/\sqrt{1-\kappa/\kappa_c}$ due to finite conductivity. The spectral
contribution to the interaction between cylinders at distance $d$ is
peaked around $\kappa\sim 1/d$. If $d \lesssim 1/\kappa_c$
($d/R\lesssim \sqrt{d/\lambda_\sigma}$, see Fig.~\ref{crossovers}(b)),
collective charge fluctuations contribute strongly to the interaction
and render it universal similar to perfect metal cylinders for which
$\kappa_c\sim 1/\sigma \to 0$. In the asymptotic regime with $d
\gtrsim 1/\kappa_c$ ($d/R\gtrsim \sqrt{d/\lambda_\sigma}$, see
Fig.~\ref{crossovers}(b)), finite conductivity prevents fluctuations on
arbitrarily large scales and hence the interaction is proportional to
$\sigma$, i.e., non-universal.  It is important to note that as $R$
goes to zero, $\kappa_c$ becomes larger, and in consequence the finite
conductivity of {the} cylinder becomes more important.

\acknowledgements We thank M.~Kardar for useful
conversations regarding this work.  This work was supported by the
NSF through grants DMR-06-45668 (RZ) and PHY0970161 (UM),  DARPA contract
No.~S-000354 (UM, RZ and TE) and DOE grant No.~DEF010204ER46131 (UM).

\begin{appendix}
%\begin{center}
%{\bf APPENDICES}
%\end{center}

\section{T-matrices}
\subsection{T-matrix of a cylinder}
\label{cyltmatrix}
In this subsection, we derive the T-matrix of a dielectric cylinder 
that is placed in the vacuum. For this purpose, in part (\ref{harmonics}) 
we find a solution to the vector wave equation in terms of 
vector cylindrical harmonics. 
In part (\ref{scattering-amplitudes}) we expand the 
electromagnetic field inside and outside of a cylinder
in the basis of the solutions presented in the previous part. Then we find the 
expansion coefficients, which are the T-matrix elements, 
for the fields inside and outside the cylinder 
by matching the boundary conditions at the cylinder surface. 
Finally in part (\ref{perfect-conductivity}) 
we show that the derived T-matrix  
in the limit of perfect conductivity  ($\epsilon \to \infty $) 
agrees with the T-matrix of a perfectly conducting cylinder. 

\subsubsection{Vector cylindrical harmonics}
\label{harmonics}
Vector cylindrical harmonics provide a basis in which divergence-less
solutions of the vector Helmholtz equation can be expanded. As in the
spherical case, the vector harmonics can be obtained by applying the
curl operator to a vector field $\Phi_{k_z n} (\kappa,{\bfx})$ that is given in terms of the
scalar harmonics. In the cylindrical case, however, this construction is
simpler than in the spherical case since the vector field $\Phi_{k_z n} (\kappa,{\bfx})$ can
be chosen to be parallel to the $z$-axis and hence does not change
direction as it does in the spherical case where $\Phi_{lm} (\kappa,{\bfx}) \sim \bfx$.
Depending on the polarization we defined the two sets of
regular cylindrical harmonics
\begin{eqnarray}
  \label{eq:vch_m}
  \bM^{\rm reg}_{k_z n}(\kappa,\bfx) & = & \frac{1}{\sqrt{k_z^2+\kappa^2}} \nabla\times \Phi^{\rm reg}_{k_z n}(\kappa,\bfx) \,,\\
  \label{eq:vch_e}
  \bN^{\rm reg}_{k_z n}(\kappa,\bfx) & = & \frac{1}{\kappa\sqrt{k_z^2+\kappa^2}}\nabla\times \nabla\times \Phi^{\rm reg}_{k_z n}(\kappa,\bfx)\,,
\end{eqnarray}
for magnetic multipoles (TE waves) and electric multipoles (TM waves),
respectively, with $p=\sqrt{k_z^2+\kappa^2}$ and the vector field
\begin{equation}
  \label{eq:V-vector-field}
  \Phi^{\rm reg}_{k_z n}(\kappa,\bfx) = \hat \bz \, I_n(\rho p) e^{in\theta} e^{ik_z z} \, ,
\end{equation}
where $\rho$, $\theta$ and $z$ are the cylindrical coordinates of $\bfx$.
The analog basis $\bM^{\rm out}_{k_z n}(\kappa,\bfx) $, $\bN^{\rm out}_{k_z n}(\kappa,\bfx) $ for outgoing waves
is obtained by replacing $I_n$ by $K_n$. These are transverse waves, i.e.,
$\nabla \bM^x_{k_z n} = \nabla \bN^x_{k_z n} =0$. They obey the relations
$\bM^x_{k_z n} = \frac{1}{i\kappa} \nabla \times \bN^x_{k_zn}$,
$\bN^x_{k_z n} = \frac{1}{i\kappa} \nabla \times \bM^x_{k_z n}$.
In explicit form, they read
\begin{gather}
\label{eq:expl_m}
\bM^{\rm reg}_{k_z n}(\kappa,\bfx)  =  \left[ \frac{in}{p\rho} I_n(p\rho) \, \hat {\bf\rho}
-I'_n(p \rho)  \hat {\bf\theta} \right] e^{in\theta} e^{ik_z z}\,, \\
\label{eq:expl_e}
\bN^{\rm reg}_{k_z n}(\kappa,\bfx)  =  \frac{1}{\kappa} \left[ 
ik_z I'_n(p\rho) \, \hat {\bf\rho}  
-\frac{nk_z}{p\rho} I_n(p\rho) \, \hat {\bf\theta} \right. \notag \\ 
\hspace{3cm}\left. - pI_n(p\rho) \hat \bz
\phantom{ \frac{1}{\kappa}}\hspace{-3mm}\right] e^{in\theta} e^{ik_z z}\,.
\end{gather}
It is analogous for the outgoing waves. 
\subsubsection{Scattering amplitudes}
\label{scattering-amplitudes}
We consider an infinitely long dielectric cylinder with $\epsilon(i c\kappa)$,
$\mu(i c\kappa)$ and radius $R$ in vacuum.  We expand the electromagnetic
field inside and outside the cylinder in the bases of
Eqs.~(\ref{eq:vch_m}), (\ref{eq:vch_e}) and the corresponding bases
for outgoing waves. The expansion coefficients for the field inside
and outside (T-matrix elements) follow from the matching conditions at
the cylinder surface for the field components that are parallel to the
surface. 

For an incident magnetic multipole (TE) field, we make the scattering ansatz
for the electric field modes

%\begin{equation}
%  \label{eq:ansatz_m_out_int}
%%  \bE_{k_z n}^M(\kappa,\bfx) = \bM^{\rm reg}_{k_z n}(\kappa,\bfx) +
%\int \frac{L dk'_z}{2\pi}\sum_{n'}[ \bM^{\rm out}_{k'_z n'}(\kappa,\bfx){\mathcal F}^{ee}_{k'_z n' M,k_z nM} + \bN^{\rm out}_{k'_z n'}(\kappa,\bfx) {\mathcal F}^{ee}_{k'_z n' M,k_z nE}  ]
%\end{equation}

%since ${\bf\mathcal F}^{ee}$ is diagonal in $k_z$ and $n$ we have
%\begin{eqnarray}
%{\mathcal F}^{ee}_{k'_z n' M,k_z nM}&=&\frac{2\pi}{L} \delta (k_z - k'_z) \delta_{n,n'} T^{MM}_{k_z n}  \nonumber \\
%{\mathcal F}^{ee}_{k'_z n' M,k_z nE}&=&\frac{2\pi}{L} \delta (k_z - k'_z) \delta_{n,n'} T^{ME}_{k_z n}  \nonumber
%\end{eqnarray}
%thus
\begin{multline}
  \label{eq:ansatz_m_out}
  \bE_{k_z n}^M(\kappa,\bfx) = \bM^{\rm reg}_{k_z n}(\kappa,\bfx) +
T^{MM}_{k_z n} \bM^{\rm out}_{k_z n}(\kappa,\bfx) \\
+  T^{ME}_{k_z n} \bN^{\rm out}_{k_z n}(\kappa,\bfx) \,,
\end{multline}
outside the cylinder and
%
%\begin{equation}
%  \label{eq:ansatz_m_in_int}
%   \bE_{k_z n}^M(\kappa,\bfx) = 
%\int \frac{L dk'_z}{2\pi}\sum_{n'}[ \tilde\bM^{\rm reg}_{k'_z n'}(\kappa,\bfx) {\mathcal F}^{ie}_{k'_z n' E,k_z nM} + \tilde\bN^{\rm reg}_{k'_z n'}(\kappa,\bfx) {\mathcal F}^{ie}_{k'_z n' E,k_z nE}  ]
%\end{equation}
%
%similar to above  ${\bf\mathcal{F}}^{ie}$ is diagonal in $k_z$ and $n$ 
%\begin{eqnarray}
%{\mathcal F}^{ie}_{k'_z n' M,k_z nM}&=&\frac{2\pi}{L} \delta (k_z - k'_z) \delta_{n,n'} A^{MM}_{k_z n}  \nonumber \\
%{\mathcal F}^{ie}_{k'_z n' M,k_z nE}&=&\frac{2\pi}{L} \delta (k_z - k'_z) \delta_{n,n'} A^{ME}_{k_z n}  \nonumber
%\end{eqnarray}
%
%therefore
%
\begin{equation}
  \label{eq:ansatz_m_in}
  \bE_{k_z n}^M(\kappa,\bfx) = A^{MM}_{k_z n} \tilde\bM^{\rm reg}_{k_z n}(\kappa,\bfx) +  A^{ME}_{k_z n} \tilde\bN^{\rm reg}_{k_z n}(\kappa,\bfx) \,,
\end{equation}
inside the cylinder where $ \tilde\bM^{\rm reg}_{k_z n}(\kappa,\bfx)$, $
\tilde\bN^{\rm reg}_{k_z n}(\kappa,\bfx)$ are given by Eqs..~(\ref{eq:vch_m}),
(\ref{eq:vch_e}) with $\kappa$ replaced by
$\kappa \sqrt{\epsilon(i c\kappa)\mu(i c\kappa)}$. For an incident electric multipole (TM) field,
the ansatz becomes
\begin{multline}
  \label{eq:ansatz_m_out}
  \bE_{k_z n}^E(\kappa,\bfx) = \bN^{\rm reg}_{k_z n}(\kappa,\bfx) +
T^{EM}_{k_z n} \bM^{\rm out}_{k_z n}(\kappa,\bfx) \\
+  T^{EE}_{k_z n} \bN^{\rm out}_{k_z n}(\kappa,\bfx) \,,
\end{multline}
outside the cylinder and
\begin{equation}
  \label{eq:ansatz_m_in}
  \bE_{k_z m}^E(\bfx) = A^{EM}_{k_z m} \tilde\bM^{\rm reg}_{k_z m}(\kappa,\bfx) +  A^{EE}_{k_z m} \tilde\bN^{\rm reg}_{k_z m}(\kappa,\bfx) \,,
\end{equation}
inside the cylinder.

The continuity conditions require that the tangential fields 
$E_z$, $E_\phi$, $H_z=B_z/\mu$
and $H_\phi=B_\phi/\mu$ are continuous across the cylinder
surface. Using the explicit expressions of Eqs.~(\ref{eq:expl_m}),
(\ref{eq:expl_e}) these conditions lead for each type of multipole
fields to a set of four linear equations for the expansion
coefficients. Using $\bB=-(1/\kappa) \nabla\times \bE$ and setting 
$p'=\sqrt{\epsilon\mu \kappa^2+k_z^2}$ these equations can be written for
incident magnetic (TE) waves as
\\
\begin{equation}
  \label{eq:lin_eqs_m}
  M_{k_z n} \begin{pmatrix} A^{MM}_{k_z n} \\ T^{MM}_{k_z n} \\ A^{ME}_{k_z n} \\ T^{ME}_{k_z n} \end{pmatrix}
= \begin{pmatrix}
\frac{p}{\kappa}  I_n(pR) \\ I'_n(pR)\\0\\ \frac{nk_z }{pR\kappa} I_n(pR)
\end{pmatrix}\,,
\end{equation}
with the matrix
\begin{widetext}
\begin{equation}
  \label{eq:m-matrix_lin_eqs}
  M_{k_z n} = \begin{pmatrix}
\frac{p'}{\mu\kappa} I_n(p'R) & -\frac{p}{\kappa} K_n(pR) & 0 & 0\\
I'_n(p' R) & -K'_n(pR) & \frac{nk_z}{p' \sqrt{\epsilon\mu} R \kappa} I_n(p'R) & 
-\frac{n k_z}{pR\kappa} K_n(pR) \\
0 & 0 & \frac{p'}{\sqrt{\epsilon\mu} \kappa} I_n(p'R) & -\frac{p}{\kappa} K_n(pR) \\
\frac{n k_z}{\mu p'R\kappa} I_n(p'R) & -\frac{n k_z}{pR\kappa} K_n(pR) & 
\sqrt{\epsilon/\mu} I'_n(p'R) & -K'_n(pR) \\
\end{pmatrix}\,,
\end{equation}
\end{widetext}

For incident electric (TM) waves the linear equations are
\begin{equation}
  \label{eq:lin_eqs_e}
  M_{k_z n} \begin{pmatrix} A^{EM}_{k_z n} \\ T^{EM}_{k_z n} \\ A^{EE}_{k_z n} \\ T^{EE}_{k_z n} \end{pmatrix}
= \begin{pmatrix}
0\\ \frac{nk_z }{pR\kappa} I_n(pR)\\\frac{p}{\kappa}  I_n(pR)\\ I'_n(pR)\\
\end{pmatrix}\,,
\end{equation}
with the same matrix $M_{k_z n}$ as before. The solution to these equations for the
T-matrix elements can be expressed as
\begin{gather}
  \label{eq:T-matrix-elements-mm}
  T^{MM}_{k_z n} = -\frac{I_n(pR)}{K_n(pR)} \frac{\Delta_1\Delta_4 +K^2}{\Delta_1\Delta_2+K^2}\,,\\
  \label{eq:T-matrix-elements-ee}
  T^{EE}_{k_z n} = -\frac{I_n(pR)}{K_n(pR)} \frac{\Delta_2\Delta_3 +K^2}{\Delta_1\Delta_2+K^2}\,,\\
  \label{eq:T-matrix-elements-me}
  T^{ME}_{k_z n} = -T^{EM}_{k_z n}=  \frac{K}{ \sqrt{\epsilon\mu} (pR)^2 K_n(pR)^2}
\frac{1}{\Delta_1\Delta_2 +K^2} \,,
\end{gather}
with
\begin{equation}
  \label{eq:def_K}
  K = \frac{n k_z}{\sqrt{\epsilon\mu} R^2 \kappa} \left( \frac{1}{p'^2} - \frac{1}{p^2}\right)\,,
\end{equation}
and
\begin{eqnarray}
  \label{eq:def_Delats}
  \Delta_1 &= & \frac{I'_n(p'R)}{p' R I_n(p'R)} -\frac{1}{\epsilon} \frac{K'_m(pR)}{pR K_m(pR)}\,,\\
 \Delta_2 &= & \frac{I'_n(p'R)}{p' R I_n(p'R)} -\frac{1}{\mu} \frac{K'_n(pR)}{pR K_n(pR)}\,,\\
 \Delta_3 &= & \frac{I'_n(p'R)}{p' R I_n(p'R)} -\frac{1}{\epsilon} \frac{I'_n(pR)}{pR I_n(pR)}\,,\\
 \Delta_4 &= & \frac{I'_n(p'R)}{p' R I_n(p'R)} -\frac{1}{\mu} \frac{I'_n(pR)}{pR I_n(pR)} \, .
\end{eqnarray}
Notice that in general the polarization is {\it not conserved} under scattering, i.e.,
$T^{EM}_{k_z n} \neq 0 \neq T^{ME}_{k_z n}$. 
\subsubsection{Limit of perfect conductivity}
\label{perfect-conductivity}
We consider the limit $\epsilon\to\infty$ with $\mu$ fixed. 
Then $p'\to \sqrt{\epsilon\mu} \,\kappa$ and $K\sim 1/\sqrt{\epsilon\mu}$. In addition, we have
for large $p'$
\begin{equation}
  \label{eq:J_m_asymp}
   \frac{I'_n(p'R)}{p' R I_n(p'R)} \to \frac{1}{p' R} \to \frac{1}{\sqrt{\epsilon\mu}R\kappa} \,,
\end{equation}
so that 
\begin{eqnarray}
  \label{eq:Deltas_asym}
  \Delta_1,\,\Delta_3 &\to& \frac{1}{\sqrt{\epsilon\mu}R \kappa} +{\cal O}(\epsilon^{-1})\,,\\
  \Delta_2 &\to& -\frac{1}{\mu} \frac{K'_n(pR)}{pR K_n(pR)} +{\cal O}(\epsilon^{-1/2})\,,\\ 
  \Delta_4 &\to& -\frac{1}{\mu}  \frac{I'_n(pR)}{pR I_n(pR)} +{\cal O}(\epsilon^{-1/2}) \, .
\end{eqnarray}
This asymptotic forms show that the T-matrix elements that couple TM and TE waves vanish
as
\begin{equation}
  \label{eq:T_me_asymp}
  T^{ME}_{k_zn}  \sim \frac{1}{\sqrt{\epsilon}} \to 0\,,
\end{equation}
in the limit $\epsilon\to\infty$. The T-matrix elements that couple like polarizations
simplify substantially. Since for $\epsilon\to\infty$
\begin{eqnarray}
  \label{eq:2}
  \frac{\Delta_1\Delta_4 -K^2}{\Delta_1\Delta_2-K^2} &\to& \frac{I_n'(pR) K_n(pR)}{I_n(pR) K'_n(pR)}\,,\\
 \frac{\Delta_2\Delta_3 -K^2}{\Delta_1\Delta_2-K^2} &\to& 1\,,
\end{eqnarray}
we get the simplified expressions
\begin{eqnarray}
  \label{eq:T_perf_metal}
  T^{MM}_{k_z n} &= & -\frac{I_n'(pR)}{ K'_n(pR)}\,, \\
  T^{EE}_{k_z n} &= & -\frac{I_n(pR)}{ K_n(pR)}\,, \\
  T^{EM}_{k_z n}&=&T^{ME}_{k_z n} = 0\, .
\end{eqnarray}
It is easily checked that these are the T-matrix elements for a scalar
field with Neumann boundary conditions (magnetic or TE modes) and with
Dirichlet boundary conditions (electric or TM modes). Hence, in the
limit of perfect conductivity the EM scattering problem for a cylinder
separates into two {\it independent} scalar problems, one with
Dirichlet and one with Neumann boundary conditions.
\subsection{T-matrix of a plate}
\label{platetmatrix}
The T-matrix elements of a plane is given by its Frensel coefficients \cite{rahi09} 
\begin{equation}
\label{tplate}
T^{P}_{{\bf k}_{\bot}}= r^P(ic \kappa,(1+{\bf k}_{\bot}^2/\kappa^2)^{-1/2})\,,
\end{equation}
with $P$ the polarization index, ${\bf k}_{\bot}$ the momentum perpendicular to the 
$\hat{\bf x}$ direction, and 
$r^Q$ the Frensel coefficients
\begin{eqnarray}
r^M(ic \kappa,x) & =
\frac
{\mu(ic \kappa) - \sqrt{1+(n^2(ic \kappa)-1)x^2}}
{\mu(ic \kappa) + \sqrt{1+(n^2(ic \kappa)-1)x^2}} \,, \\
r^E(ic \kappa,x) & =
\frac
{\epsilon(ic \kappa) - \sqrt{1+(n^2(ic \kappa)-1)x^2}}
{\epsilon(ic \kappa) + \sqrt{1+(n^2(ic \kappa)-1)x^2}},
\label{Fresnel}
\end{eqnarray}
here $\epsilon$, $\mu$ and  $n(ic \kappa)$ are the dielectric response function, magnetic permeability,  and the refractive index of the plate, respectively. The refractive index in terms of the dielectric function and the magnetic permeability of the plate is given by
\begin{equation}
 n(ic \kappa)=\sqrt{\epsilon(ic \kappa)\mu(ic \kappa)}\,.
 \end{equation}
The T-matrix of a dielectric plate Eq.~(\ref{tplate}) is diagonal with respect to the polarization indices.

\section{Translation matrix}
\label{translation-matrix}

According to Graf's addition theorem, the following relation holds,
\begin{multline}
  \label{eq:add_theorem}
  K_m(p\,r_j)\, e^{im\phi_j} = \\ \sum_{n=-\infty}^\infty (-1)^n K_{m-n}(p\,R_{ji})
\,e^{-i(m-n)\,\phi_{ji}} I_m(p\,r_i)\, e^{in\phi_i} \, ,
\end{multline}
where $\br_j=\br_i+\bR_{ji}$ with $\br_i=r_i(\cos\phi_i,\sin\phi_i)$,
$\bR_{ji} = R_{ji}(\cos \phi_{ji},\sin \phi_{ji})$ are 
two-dimensional vectors in the $xy$-plane. We consider translations in 3D that are perpendicular
to the $z$-axis with the translation vector $\bX_{ji}=(\bR_{ji},0)$, i.e., we
set $\bfx_i=(\br_i,z_i)$ with $z_i=z_j$. Since the curl operator commutes
with translations, from the definitions in Eqs.~(\ref{eq:vch_m}), (\ref{eq:vch_e}) and
the addition theorem of Eq.~(\ref{eq:add_theorem}) follow the translation formulas from outgoing
to regular waves
\begin{eqnarray}
  \label{eq:trans_M}
  \bM^{\rm out}_{k_z m}(\bfx_j) & = &  \sum_{n=-\infty}^\infty {\cU}^{ij}_{k_z nm} \, \bM^{\rm reg}_{k_z n}(\bfx_i) \\ 
  \label{eq:trans_N}
  \bN^{\rm out}_{k_z m}(\bfx_j) & = &  \sum_{n=-\infty}^\infty {\cU}^{ij}_{k_z nm} \, \bN^{\rm reg}_{k_z n}(\bfx_i) 
\end{eqnarray}
with the translation matrix
\begin{equation}
  \label{eq:B-matrix}
  \cU^{ij}_{k_z nm} = K_{m-n} (p| \bX_{ji} |) e^{-i(m-n)\phi_{ji}} \, .
\end{equation}
From this we make two important observations: Translations conserve the polarization, i.e.,
they do not couple magnetic and electric modes, and the translation matrices are diagonal
in $k_z$. The conservation of polarization leads to a diagonal translation matrix 
that acts on the full set of electric and magnetic modes.

\section{Conversion matrix from vector plane wave basis to cylindrical vector wave basis}
\label{conv}
In this section we show the conversion matrix elements from Ref.~\cite{rahi09}. 
The cylindrical vector wave functions are given in Eq.~\eqref{eq:vch_m} and \eqref{eq:vch_e}, 
 which decay along the $-\mathbf{\hat{z}}$.  
We consider regular vector plane wave functions
that decay along the $-\mathbf{\hat{x}}$ axis 
\begin{multline}
{{\bf M}^\text{reg}}_{\veckpe}(\kappa,\vecx)  = \\ \frac{1}{\sqrt{k_y^2+k_z^2}}\curl
e^{\sqrt{\kappa^2+k_y^2+k_z^2} x + i k_y y + i k_z z} \mathbf{\hat{x}},
\end{multline}
\begin{multline}
{{\bf N}^\text{reg}}_{\veckpe}(\kappa,\vecx) = \\ \frac{1}{\kappa\sqrt{k_y^2+k_z^2}} \curl
\curl e^{\sqrt{\kappa^2+k_y^2+k_z^2} x + i k_y y + i k_z z} \mathbf{\hat{x}}.
\end{multline}
The vector plane wave functions can be written in terms of vector
cylindrical wave functions,
\begin{gather}
\label{convMNcylinderplane}
{{\bf M}^\text{reg}}_{\veckpe}(\kappa,\vecx)  = \sum_{n} D_{k_z n M,\veckpe M}
{{\bf M}^\text{reg}}_{k_z n}(\kappa,\vecx)\notag \\ + D_{k_z n E,\veckpe M} 
{{\bf N}^\text{reg}}_{k_z n}(\kappa,\vecx) \, , \\
{{\bf N}^\text{reg}}_{\veckpe}(\kappa,\vecx)  = \sum_{n} D_{k_z n M,\veckpe E}
{{\bf M}^\text{reg}}_{k_z n}(\kappa,\vecx) \notag\\+ D_{k_z n E,\veckpe E} 
{{\bf N}^\text{reg}}_{k_z n}(\kappa,\vecx) \, ,
\end{gather}
using the conversion matrix elements
\begin{align}
D_{k_z n M, \veckpe M} &  = -i \frac{k_z}{\sqrt{k_y^2+k_z^2}}
\sqrt{1+\xi^2} \left(\sqrt{1+\xi^2}+\xi\right)^n \, ,\\
D_{k_z n E, \veckpe M} & = i \frac{\kappa}{\sqrt{k_y^2+k_z^2}}
\xi \left(\sqrt{1+\xi^2}+\xi\right)^n \, , \\
D_{k_z n E, \veckpe E} & = D_{k_z n M, \veckpe M} \, ,\\
D_{k_z n M, \veckpe E} & = -D_{k_z n E, \veckpe M},
\end{align}
where $\xi = \frac{k_y}{\sqrt{\kappa^2+k_z^2}}$ and $\veckpe=(k_y,k_z)$.

\end{appendix}

\bibliography{refs.bib}

%merlin.mbs 2010-03-15 4.21a (PWD, AO, DPC)
\begin{thebibliography}{30}%
\makeatletter
\providecommand \@ifxundefined [1]{%
 \@ifx{#1\undefined}
}%
\providecommand \@ifnum [1]{%
 \ifnum #1\expandafter \@firstoftwo
 \else \expandafter \@secondoftwo
 \fi
}%
\providecommand \@ifx [1]{%
 \ifx #1\expandafter \@firstoftwo
 \else \expandafter \@secondoftwo
 \fi
}%
\providecommand \natexlab [1]{#1}%
\providecommand \enquote  [1]{``#1''}%
\providecommand \bibnamefont  [1]{#1}%
\providecommand \bibfnamefont [1]{#1}%
\providecommand \citenamefont [1]{#1}%
\providecommand \href@noop [0]{\@secondoftwo}%
\providecommand \href [0]{\begingroup \@sanitize@url \@href}%
\providecommand \@href[1]{\@@startlink{#1}\@@href}%
\providecommand \@@href[1]{\endgroup#1\@@endlink}%
\providecommand \@sanitize@url [0]{\catcode `\\12\catcode `\$12\catcode
  `\&12\catcode `\#12\catcode `\^12\catcode `\_12\catcode `\%12\relax}%
\providecommand \@@startlink[1]{}%
\providecommand \@@endlink[0]{}%
\providecommand \url  [0]{\begingroup\@sanitize@url \@url }%
\providecommand \@url [1]{\endgroup\@href {#1}{\urlprefix }}%
\providecommand \urlprefix  [0]{URL }%
\providecommand \Eprint [0]{\href }%
\@ifxundefined \urlstyle {%
  \providecommand \doi  [0]{\begingroup \@sanitize@url \@doi}%
  \providecommand \@doi [1]{\endgroup \@@startlink {\doibase
  #1}doi:\discretionary {}{}{}#1\@@endlink }%
}{%
  \providecommand \doi  [0]{doi:\discretionary{}{}{}\begingroup
  \urlstyle{rm}\Url }%
}%
\providecommand \doibase [0]{http://dx.doi.org/}%
\providecommand \Doi [0]{\begingroup \@sanitize@url \@Doi }%
\providecommand \@Doi  [1]{\endgroup\@@startlink{\doibase#1}\@@Doi}%
\providecommand \@@Doi [1]{#1\@@endlink}%
\providecommand \selectlanguage [0]{\@gobble}%
\providecommand \bibinfo  [0]{\@secondoftwo}%
\providecommand \bibfield  [0]{\@secondoftwo}%
\providecommand \translation [1]{[#1]}%
\providecommand \BibitemOpen [0]{}%
\providecommand \bibitemStop [0]{}%
\providecommand \bibitemNoStop [0]{.\EOS\space}%
\providecommand \EOS [0]{\spacefactor3000\relax}%
\providecommand \BibitemShut  [1]{\csname bibitem#1\endcsname}%
%</preamble>
\bibitem [{\citenamefont {Brown-Hayes}\ \emph {et~al.}(2005)\citenamefont
  {Brown-Hayes}, \citenamefont {Dalvit}, \citenamefont {Mazzitelli},
  \citenamefont {Kim},\ and\ \citenamefont {Onofrio}}]{brown05}%
  \BibitemOpen
  \bibfield  {author} {\bibinfo {author} {\bibfnamefont {M.}~\bibnamefont
  {Brown-Hayes}}, \bibinfo {author} {\bibfnamefont {D.~A.~R.}\ \bibnamefont
  {Dalvit}}, \bibinfo {author} {\bibfnamefont {F.~D.}\ \bibnamefont
  {Mazzitelli}}, \bibinfo {author} {\bibfnamefont {W.~J.}\ \bibnamefont {Kim}},
  \ and\ \bibinfo {author} {\bibfnamefont {R.}~\bibnamefont {Onofrio}},\ }\Doi
  {10.1103/PhysRevA.72.052102} {\bibfield  {journal} {\bibinfo  {journal}
  {Phys. Rev. A},\ }\textbf {\bibinfo {volume} {72}},\ \bibinfo {pages}
  {052102} (\bibinfo {year} {2005})}\BibitemShut {NoStop}%
\bibitem [{\citenamefont {Decca}\ \emph {et~al.}(2010)\citenamefont {Decca},
  \citenamefont {Fischbach}, \citenamefont {Klimchitskaya}, \citenamefont
  {Krause}, \citenamefont {L\'opez},\ and\ \citenamefont
  {Mostepanenko}}]{decca10}%
  \BibitemOpen
  \bibfield  {author} {\bibinfo {author} {\bibfnamefont {R.~S.}\ \bibnamefont
  {Decca}}, \bibinfo {author} {\bibfnamefont {E.}~\bibnamefont {Fischbach}},
  \bibinfo {author} {\bibfnamefont {G.~L.}\ \bibnamefont {Klimchitskaya}},
  \bibinfo {author} {\bibfnamefont {D.~E.}\ \bibnamefont {Krause}}, \bibinfo
  {author} {\bibfnamefont {D.}~\bibnamefont {L\'opez}}, \ and\ \bibinfo
  {author} {\bibfnamefont {V.~M.}\ \bibnamefont {Mostepanenko}},\ }\Doi
  {10.1103/PhysRevA.82.052515} {\bibfield  {journal} {\bibinfo  {journal}
  {Phys. Rev. A},\ }\textbf {\bibinfo {volume} {82}},\ \bibinfo {pages}
  {052515} (\bibinfo {year} {2010})}\BibitemShut {NoStop}%
\bibitem [{\citenamefont {Sazonova}\ \emph {et~al.}(2004)\citenamefont
  {Sazonova}, \citenamefont {Yaish}, \citenamefont {Ustunel}, \citenamefont
  {Roundy}, \citenamefont {Arias},\ and\ \citenamefont {Mceuen}}]{saz04}%
  \BibitemOpen
  \bibfield  {author} {\bibinfo {author} {\bibfnamefont {V.}~\bibnamefont
  {Sazonova}}, \bibinfo {author} {\bibfnamefont {Y.}~\bibnamefont {Yaish}},
  \bibinfo {author} {\bibfnamefont {H.}~\bibnamefont {Ustunel}}, \bibinfo
  {author} {\bibfnamefont {D.}~\bibnamefont {Roundy}}, \bibinfo {author}
  {\bibfnamefont {T.~A.}\ \bibnamefont {Arias}}, \ and\ \bibinfo {author}
  {\bibfnamefont {P.~L.}\ \bibnamefont {Mceuen}},\ }\Doi {10.1038/nature02905}
  {\bibfield  {journal} {\bibinfo  {journal} {Nature (London)},\ }\textbf
  {\bibinfo {volume} {431}},\ \bibinfo {pages} {284} (\bibinfo {year}
  {2004})}\BibitemShut {NoStop}%
\bibitem [{\citenamefont {Serry}\ \emph {et~al.}(1998)\citenamefont {Serry},
  \citenamefont {Walliser},\ and\ \citenamefont {Maclay}}]{serry98}%
  \BibitemOpen
  \bibfield  {author} {\bibinfo {author} {\bibfnamefont {F.~M.}\ \bibnamefont
  {Serry}}, \bibinfo {author} {\bibfnamefont {D.}~\bibnamefont {Walliser}}, \
  and\ \bibinfo {author} {\bibfnamefont {G.~J.}\ \bibnamefont {Maclay}},\ }\Doi
  {10.1063/1.368410} {\bibfield  {journal} {\bibinfo  {journal} {Journal of
  Applied Physics},\ }\textbf {\bibinfo {volume} {84}},\ \bibinfo {pages} {2501
  } (\bibinfo {year} {1998})},\ ISSN \bibinfo {issn} {0021-8979}\BibitemShut
  {NoStop}%
\bibitem [{\citenamefont {Buks}\ and\ \citenamefont {Roukes}(2001)}]{buk01}%
  \BibitemOpen
  \bibfield  {author} {\bibinfo {author} {\bibfnamefont {E.}~\bibnamefont
  {Buks}}\ and\ \bibinfo {author} {\bibfnamefont {M.~L.}\ \bibnamefont
  {Roukes}},\ }\Doi {10.1103/PhysRevB.63.033402} {\bibfield  {journal}
  {\bibinfo  {journal} {Phys. Rev. B},\ }\textbf {\bibinfo {volume} {63}},\
  \bibinfo {pages} {033402} (\bibinfo {year} {2001})}\BibitemShut {NoStop}%
\bibitem [{\citenamefont {\ifmmode~\check{S}\else \v{S}\fi{}iber}\ \emph
  {et~al.}(2009)\citenamefont {\ifmmode~\check{S}\else \v{S}\fi{}iber},
  \citenamefont {Rajter}, \citenamefont {French}, \citenamefont {Ching},
  \citenamefont {Parsegian},\ and\ \citenamefont {Podgornik}}]{podgo}%
  \BibitemOpen
  \bibfield  {author} {\bibinfo {author} {\bibfnamefont {A.}~\bibnamefont
  {\ifmmode~\check{S}\else \v{S}\fi{}iber}}, \bibinfo {author} {\bibfnamefont
  {R.~F.}\ \bibnamefont {Rajter}}, \bibinfo {author} {\bibfnamefont {R.~H.}\
  \bibnamefont {French}}, \bibinfo {author} {\bibfnamefont {W.~Y.}\
  \bibnamefont {Ching}}, \bibinfo {author} {\bibfnamefont {V.~A.}\ \bibnamefont
  {Parsegian}}, \ and\ \bibinfo {author} {\bibfnamefont {R.}~\bibnamefont
  {Podgornik}},\ }\Doi {10.1103/PhysRevB.80.165414} {\bibfield  {journal}
  {\bibinfo  {journal} {Phys. Rev. B},\ }\textbf {\bibinfo {volume} {80}},\
  \bibinfo {pages} {165414} (\bibinfo {year} {2009})}\BibitemShut {NoStop}%
\bibitem [{\citenamefont {Elliott}(1968)}]{gf68}%
  \BibitemOpen
  \bibfield  {author} {\bibinfo {author} {\bibfnamefont {G.~F.}\ \bibnamefont
  {Elliott}},\ }\Doi {10.1016/0022-5193(68)90060-X} {\bibfield  {journal}
  {\bibinfo  {journal} {Journal of Theoretical Biology},\ }\textbf {\bibinfo
  {volume} {21}},\ \bibinfo {pages} {71 } (\bibinfo {year} {1968})},\ ISSN
  \bibinfo {issn} {0022-5193}\BibitemShut {NoStop}%
\bibitem [{\citenamefont {Parsegian}(1972)}]{parseg72}%
  \BibitemOpen
  \bibfield  {author} {\bibinfo {author} {\bibfnamefont {V.~A.}\ \bibnamefont
  {Parsegian}},\ }\Doi {10.1063/1.1677878} {\bibfield  {journal} {\bibinfo
  {journal} {J. Chem. Phys.},\ }\textbf {\bibinfo {volume} {56}} (\bibinfo
  {year} {1972})},\ ISSN \bibinfo {issn} {0021-9606},\ \doi
  {10.1063/1.1677878}\BibitemShut {NoStop}%
\bibitem [{\citenamefont {Emig}\ \emph {et~al.}(2006)\citenamefont {Emig},
  \citenamefont {Jaffe}, \citenamefont {Kardar},\ and\ \citenamefont
  {Scardicchio}}]{emig06}%
  \BibitemOpen
  \bibfield  {author} {\bibinfo {author} {\bibfnamefont {T.}~\bibnamefont
  {Emig}}, \bibinfo {author} {\bibfnamefont {R.~L.}\ \bibnamefont {Jaffe}},
  \bibinfo {author} {\bibfnamefont {M.}~\bibnamefont {Kardar}}, \ and\ \bibinfo
  {author} {\bibfnamefont {A.}~\bibnamefont {Scardicchio}},\ }\Doi
  {10.1103/PhysRevLett.96.080403} {\bibfield  {journal} {\bibinfo  {journal}
  {Phys. Rev. Lett.},\ }\textbf {\bibinfo {volume} {96}},\ \bibinfo {pages}
  {080403} (\bibinfo {year} {2006})}\BibitemShut {NoStop}%
\bibitem [{\citenamefont {Rahi}\ \emph {et~al.}(2008)\citenamefont {Rahi},
  \citenamefont {Emig}, \citenamefont {Jaffe},\ and\ \citenamefont
  {Kardar}}]{rahi08}%
  \BibitemOpen
  \bibfield  {author} {\bibinfo {author} {\bibfnamefont {S.~J.}\ \bibnamefont
  {Rahi}}, \bibinfo {author} {\bibfnamefont {T.}~\bibnamefont {Emig}}, \bibinfo
  {author} {\bibfnamefont {R.~L.}\ \bibnamefont {Jaffe}}, \ and\ \bibinfo
  {author} {\bibfnamefont {M.}~\bibnamefont {Kardar}},\ }\Doi
  {10.1103/PhysRevA.78.012104} {\bibfield  {journal} {\bibinfo  {journal}
  {Phys. Rev. A},\ }\textbf {\bibinfo {volume} {78}},\ \bibinfo {pages}
  {012104} (\bibinfo {year} {2008})}\BibitemShut {NoStop}%
\bibitem [{\citenamefont {Rahi}\ \emph {et~al.}(2009)\citenamefont {Rahi},
  \citenamefont {Emig}, \citenamefont {Graham}, \citenamefont {Jaffe},\ and\
  \citenamefont {Kardar}}]{rahi09}%
  \BibitemOpen
  \bibfield  {author} {\bibinfo {author} {\bibfnamefont {S.~J.}\ \bibnamefont
  {Rahi}}, \bibinfo {author} {\bibfnamefont {T.}~\bibnamefont {Emig}}, \bibinfo
  {author} {\bibfnamefont {N.}~\bibnamefont {Graham}}, \bibinfo {author}
  {\bibfnamefont {R.~L.}\ \bibnamefont {Jaffe}}, \ and\ \bibinfo {author}
  {\bibfnamefont {M.}~\bibnamefont {Kardar}},\ }\Doi
  {10.1103/PhysRevD.80.085021} {\bibfield  {journal} {\bibinfo  {journal}
  {Phys. Rev. D},\ }\textbf {\bibinfo {volume} {80}},\ \bibinfo {pages}
  {085021} (\bibinfo {year} {2009})}\BibitemShut {NoStop}%
\bibitem [{\citenamefont {U.~Mohideen}\ and\ \citenamefont
  {Mostepanenko}(2009)}]{umar}%
  \BibitemOpen
  \bibfield  {author} {\bibinfo {author} {\bibfnamefont {M.~B.}\ \bibnamefont
  {U.~Mohideen}, \bibfnamefont {G.~L.~Klimchitskaya}}\ and\ \bibinfo {author}
  {\bibfnamefont {V.~M.}\ \bibnamefont {Mostepanenko}},\ }\href@noop {} {\emph
  {\bibinfo {title} {Advances in the Casimir Effect}}}\ (\bibinfo  {publisher}
  {Oxford University Press},\ \bibinfo {year} {2009})\BibitemShut {NoStop}%
\bibitem [{\citenamefont {Emig}\ \emph {et~al.}(2009)\citenamefont {Emig},
  \citenamefont {Graham}, \citenamefont {Jaffe},\ and\ \citenamefont
  {Kardar}}]{Emig:2009fk}%
  \BibitemOpen
  \bibfield  {author} {\bibinfo {author} {\bibfnamefont {T.}~\bibnamefont
  {Emig}}, \bibinfo {author} {\bibfnamefont {N.}~\bibnamefont {Graham}},
  \bibinfo {author} {\bibfnamefont {R.~L.}\ \bibnamefont {Jaffe}}, \ and\
  \bibinfo {author} {\bibfnamefont {M.}~\bibnamefont {Kardar}},\ }\href@noop {}
  {\bibfield  {journal} {\bibinfo  {journal} {Phys. Rev. A},\ }\textbf
  {\bibinfo {volume} {79}},\ \bibinfo {pages} {054901} (\bibinfo {year}
  {2009})}\BibitemShut {NoStop}%
\bibitem [{\citenamefont {Graham}\ \emph {et~al.}(2010)\citenamefont {Graham},
  \citenamefont {Shpunt}, \citenamefont {Emig}, \citenamefont {Rahi},
  \citenamefont {Jaffe},\ and\ \citenamefont {Kardar}}]{Graham:2010uq}%
  \BibitemOpen
  \bibfield  {author} {\bibinfo {author} {\bibfnamefont {N.}~\bibnamefont
  {Graham}}, \bibinfo {author} {\bibfnamefont {A.}~\bibnamefont {Shpunt}},
  \bibinfo {author} {\bibfnamefont {T.}~\bibnamefont {Emig}}, \bibinfo {author}
  {\bibfnamefont {S.~J.}\ \bibnamefont {Rahi}}, \bibinfo {author}
  {\bibfnamefont {R.~L.}\ \bibnamefont {Jaffe}}, \ and\ \bibinfo {author}
  {\bibfnamefont {M.}~\bibnamefont {Kardar}},\ }\Doi
  {10.1103/PhysRevD.81.061701} {\bibfield  {journal} {\bibinfo  {journal}
  {Phys. Rev. D},\ }\textbf {\bibinfo {volume} {81}},\ \bibinfo {pages}
  {061701} (\bibinfo {year} {2010})}\BibitemShut {NoStop}%
\bibitem [{\citenamefont {Maghrebi}\ \emph {et~al.}(2011)\citenamefont
  {Maghrebi}, \citenamefont {Rahi}, \citenamefont {Emig}, \citenamefont
  {Graham}, \citenamefont {Jaffe},\ and\ \citenamefont
  {Kardar}}]{Maghrebi:2011kx}%
  \BibitemOpen
  \bibfield  {author} {\bibinfo {author} {\bibfnamefont {M.~F.}\ \bibnamefont
  {Maghrebi}}, \bibinfo {author} {\bibfnamefont {S.~J.}\ \bibnamefont {Rahi}},
  \bibinfo {author} {\bibfnamefont {T.}~\bibnamefont {Emig}}, \bibinfo {author}
  {\bibfnamefont {N.}~\bibnamefont {Graham}}, \bibinfo {author} {\bibfnamefont
  {R.~L.}\ \bibnamefont {Jaffe}}, \ and\ \bibinfo {author} {\bibfnamefont
  {M.}~\bibnamefont {Kardar}},\ }\href@noop {} {\bibfield  {journal} {\bibinfo
  {journal} {PNAS},\ }\textbf {\bibinfo {volume} {108}},\ \bibinfo {pages}
  {6867} (\bibinfo {year} {2011})}\BibitemShut {NoStop}%
\bibitem [{\citenamefont {Zandi}\ \emph {et~al.}(2010)\citenamefont {Zandi},
  \citenamefont {Emig},\ and\ \citenamefont {Mohideen}}]{zandi10}%
  \BibitemOpen
  \bibfield  {author} {\bibinfo {author} {\bibfnamefont {R.}~\bibnamefont
  {Zandi}}, \bibinfo {author} {\bibfnamefont {T.}~\bibnamefont {Emig}}, \ and\
  \bibinfo {author} {\bibfnamefont {U.}~\bibnamefont {Mohideen}},\ }\Doi
  {10.1103/PhysRevB.81.195423} {\bibfield  {journal} {\bibinfo  {journal}
  {Phys. Rev. B},\ }\textbf {\bibinfo {volume} {81}},\ \bibinfo {pages}
  {195423} (\bibinfo {year} {2010})}\BibitemShut {NoStop}%
\bibitem [{\citenamefont {Noruzifar}\ \emph {et~al.}(2011)\citenamefont
  {Noruzifar}, \citenamefont {Emig},\ and\ \citenamefont {Zandi}}]{noru11}%
  \BibitemOpen
  \bibfield  {author} {\bibinfo {author} {\bibfnamefont {E.}~\bibnamefont
  {Noruzifar}}, \bibinfo {author} {\bibfnamefont {T.}~\bibnamefont {Emig}}, \
  and\ \bibinfo {author} {\bibfnamefont {R.}~\bibnamefont {Zandi}},\ }\Doi
  {10.1103/PhysRevA.84.042501} {\bibfield  {journal} {\bibinfo  {journal}
  {Phys. Rev. A},\ }\textbf {\bibinfo {volume} {84}},\ \bibinfo {pages}
  {042501} (\bibinfo {year} {2011})}\BibitemShut {NoStop}%
\bibitem [{\citenamefont {Bordag}\ \emph {et~al.}(2006)\citenamefont {Bordag},
  \citenamefont {Geyer}, \citenamefont {Klimchitskaya},\ and\ \citenamefont
  {Mostepanenko}}]{most06}%
  \BibitemOpen
  \bibfield  {author} {\bibinfo {author} {\bibfnamefont {M.}~\bibnamefont
  {Bordag}}, \bibinfo {author} {\bibfnamefont {B.}~\bibnamefont {Geyer}},
  \bibinfo {author} {\bibfnamefont {G.}~\bibnamefont {Klimchitskaya}}, \ and\
  \bibinfo {author} {\bibfnamefont {V.}~\bibnamefont {Mostepanenko}},\
  }\href@noop {} {\bibfield  {journal} {\bibinfo  {journal} {Phys. Rev. B},\
  }\textbf {\bibinfo {volume} {74}},\ \bibinfo {pages} {205431} (\bibinfo
  {year} {2006})}\BibitemShut {NoStop}%
\bibitem [{\citenamefont {Bordag}(2006)}]{bordag06}%
  \BibitemOpen
  \bibfield  {author} {\bibinfo {author} {\bibfnamefont {M.}~\bibnamefont
  {Bordag}},\ }\Doi {10.1103/PhysRevD.73.125018} {\bibfield  {journal}
  {\bibinfo  {journal} {Phys. Rev. D},\ }\textbf {\bibinfo {volume} {73}},\
  \bibinfo {pages} {125018} (\bibinfo {year} {2006})}\BibitemShut {NoStop}%
\bibitem [{\citenamefont {Lombardo}\ \emph {et~al.}(2008)\citenamefont
  {Lombardo}, \citenamefont {Mazzitelli},\ and\ \citenamefont
  {Villar}}]{lombardo08}%
  \BibitemOpen
  \bibfield  {author} {\bibinfo {author} {\bibfnamefont {F.~C.}\ \bibnamefont
  {Lombardo}}, \bibinfo {author} {\bibfnamefont {F.~D.}\ \bibnamefont
  {Mazzitelli}}, \ and\ \bibinfo {author} {\bibfnamefont {P.~I.}\ \bibnamefont
  {Villar}},\ }\Doi {10.1103/PhysRevD.78.085009} {\bibfield  {journal}
  {\bibinfo  {journal} {Phys. Rev. D},\ }\textbf {\bibinfo {volume} {78}},\
  \bibinfo {pages} {085009} (\bibinfo {year} {2008})}\BibitemShut {NoStop}%
\bibitem [{\citenamefont {Weber}\ and\ \citenamefont {Gies}(2010)}]{holger10}%
  \BibitemOpen
  \bibfield  {author} {\bibinfo {author} {\bibfnamefont {A.}~\bibnamefont
  {Weber}}\ and\ \bibinfo {author} {\bibfnamefont {H.}~\bibnamefont {Gies}},\
  }\Doi {10.1103/PhysRevD.82.125019} {\bibfield  {journal} {\bibinfo  {journal}
  {Phys. Rev. D},\ }\textbf {\bibinfo {volume} {82}},\ \bibinfo {pages}
  {125019} (\bibinfo {year} {2010})}\BibitemShut {NoStop}%
\bibitem [{\citenamefont {Barash}\ and\ \citenamefont
  {Kyasov}(1989)}]{barash89}%
  \BibitemOpen
  \bibfield  {author} {\bibinfo {author} {\bibfnamefont {Y.}~\bibnamefont
  {Barash}}\ and\ \bibinfo {author} {\bibfnamefont {A.}~\bibnamefont
  {Kyasov}},\ }\href@noop {} {\bibfield  {journal} {\bibinfo  {journal} {Soviet
  Physics - JETP},\ }\textbf {\bibinfo {volume} {68}},\ \bibinfo {pages} {39}
  (\bibinfo {year} {1989})}\BibitemShut {NoStop}%
\bibitem [{\citenamefont {Dobson}\ \emph {et~al.}(2006)\citenamefont {Dobson},
  \citenamefont {White},\ and\ \citenamefont {Rubio}}]{dobson06}%
  \BibitemOpen
  \bibfield  {author} {\bibinfo {author} {\bibfnamefont {J.~F.}\ \bibnamefont
  {Dobson}}, \bibinfo {author} {\bibfnamefont {A.}~\bibnamefont {White}}, \
  and\ \bibinfo {author} {\bibfnamefont {A.}~\bibnamefont {Rubio}},\ }\Doi
  {10.1103/PhysRevLett.96.073201} {\bibfield  {journal} {\bibinfo  {journal}
  {Phys. Rev. Lett.},\ }\textbf {\bibinfo {volume} {96}},\ \bibinfo {pages}
  {073201} (\bibinfo {year} {2006})}\BibitemShut {NoStop}%
\bibitem [{\citenamefont {Mazzitelli}\ \emph {et~al.}(2006)\citenamefont
  {Mazzitelli}, \citenamefont {Dalvit},\ and\ \citenamefont
  {Lombardo}}]{dalvit06}%
  \BibitemOpen
  \bibfield  {author} {\bibinfo {author} {\bibfnamefont {F.~D.}\ \bibnamefont
  {Mazzitelli}}, \bibinfo {author} {\bibfnamefont {D.~A.~R.}\ \bibnamefont
  {Dalvit}}, \ and\ \bibinfo {author} {\bibfnamefont {F.~C.}\ \bibnamefont
  {Lombardo}},\ }\href@noop {} {\bibfield  {journal} {\bibinfo  {journal} {New
  Journal of Physics},\ }\textbf {\bibinfo {volume} {8}},\ \bibinfo {pages}
  {240} (\bibinfo {year} {2006})}\BibitemShut {NoStop}%
\bibitem [{\citenamefont {Drummond}\ and\ \citenamefont
  {Needs}(2007)}]{drummond07}%
  \BibitemOpen
  \bibfield  {author} {\bibinfo {author} {\bibfnamefont {N.~D.}\ \bibnamefont
  {Drummond}}\ and\ \bibinfo {author} {\bibfnamefont {R.~J.}\ \bibnamefont
  {Needs}},\ }\Doi {10.1103/PhysRevLett.99.166401} {\bibfield  {journal}
  {\bibinfo  {journal} {Phys. Rev. Lett.},\ }\textbf {\bibinfo {volume} {99}},\
  \bibinfo {pages} {166401} (\bibinfo {year} {2007})}\BibitemShut {NoStop}%
\bibitem [{\citenamefont {Dobson}\ \emph {et~al.}(2009)\citenamefont {Dobson},
  \citenamefont {Gould},\ and\ \citenamefont {Klich}}]{dobson09}%
  \BibitemOpen
  \bibfield  {author} {\bibinfo {author} {\bibfnamefont {J.~F.}\ \bibnamefont
  {Dobson}}, \bibinfo {author} {\bibfnamefont {T.}~\bibnamefont {Gould}}, \
  and\ \bibinfo {author} {\bibfnamefont {I.}~\bibnamefont {Klich}},\ }\Doi
  {10.1103/PhysRevA.80.012506} {\bibfield  {journal} {\bibinfo  {journal}
  {Phys. Rev. A},\ }\textbf {\bibinfo {volume} {80}},\ \bibinfo {pages}
  {012506} (\bibinfo {year} {2009})}\BibitemShut {NoStop}%
\bibitem [{Note1()}]{Note1}%
  \BibitemOpen
  \bibinfo {note} {In the cylinder-plate case, we restricted the numerics to
  $h/R>0.3$ with $n_{\protect \rm max}=41$ due to increasing numerical
  uncertainties in the $k_y$ integration in Eq.~(\ref {Mmatrix})}\BibitemShut
  {NoStop}%
\bibitem [{\citenamefont {Anderson}\ \emph {et~al.}(1999)\citenamefont
  {Anderson}, \citenamefont {Bai}, \citenamefont {Bischof}, \citenamefont
  {Blackford}, \citenamefont {Demmel}, \citenamefont {Dongarra}, \citenamefont
  {Du~Croz}, \citenamefont {Greenbaum}, \citenamefont {Hammarling},
  \citenamefont {McKenney},\ and\ \citenamefont {Sorensen}}]{laug}%
  \BibitemOpen
  \bibfield  {author} {\bibinfo {author} {\bibfnamefont {E.}~\bibnamefont
  {Anderson}}, \bibinfo {author} {\bibfnamefont {Z.}~\bibnamefont {Bai}},
  \bibinfo {author} {\bibfnamefont {C.}~\bibnamefont {Bischof}}, \bibinfo
  {author} {\bibfnamefont {S.}~\bibnamefont {Blackford}}, \bibinfo {author}
  {\bibfnamefont {J.}~\bibnamefont {Demmel}}, \bibinfo {author} {\bibfnamefont
  {J.}~\bibnamefont {Dongarra}}, \bibinfo {author} {\bibfnamefont
  {J.}~\bibnamefont {Du~Croz}}, \bibinfo {author} {\bibfnamefont
  {A.}~\bibnamefont {Greenbaum}}, \bibinfo {author} {\bibfnamefont
  {S.}~\bibnamefont {Hammarling}}, \bibinfo {author} {\bibfnamefont
  {A.}~\bibnamefont {McKenney}}, \ and\ \bibinfo {author} {\bibfnamefont
  {D.}~\bibnamefont {Sorensen}},\ }\href@noop {} {\emph {\bibinfo {title}
  {{LAPACK} Users' Guide}}},\ \bibinfo {edition} {3rd}\ ed.\ (\bibinfo
  {publisher} {Society for Industrial and Applied Mathematics},\ \bibinfo
  {address} {Philadelphia, PA},\ \bibinfo {year} {1999})\ ISBN \bibinfo {isbn}
  {0-89871-447-8 (paperback)}\BibitemShut {NoStop}%
\bibitem [{\citenamefont {Lifshitz}(1956)}]{Lifshitz56}%
  \BibitemOpen
  \bibfield  {author} {\bibinfo {author} {\bibfnamefont {E.~M.}\ \bibnamefont
  {Lifshitz}},\ }\href@noop {} {\bibfield  {journal} {\bibinfo  {journal} {Sov.
  Phys. JETP},\ }\textbf {\bibinfo {volume} {2}},\ \bibinfo {pages} {73}
  (\bibinfo {year} {1956})}\BibitemShut {NoStop}%
\bibitem [{\citenamefont {Decca}\ \emph {et~al.}(2007)\citenamefont {Decca},
  \citenamefont {L\'{o}pez}, \citenamefont {Fischbach}, \citenamefont
  {Klimchitskaya}, \citenamefont {Krause},\ and\ \citenamefont
  {Mostepanenko}}]{Decca07}%
  \BibitemOpen
  \bibfield  {author} {\bibinfo {author} {\bibfnamefont {R.~S.}\ \bibnamefont
  {Decca}}, \bibinfo {author} {\bibfnamefont {D.}~\bibnamefont {L\'{o}pez}},
  \bibinfo {author} {\bibfnamefont {E.}~\bibnamefont {Fischbach}}, \bibinfo
  {author} {\bibfnamefont {G.~L.}\ \bibnamefont {Klimchitskaya}}, \bibinfo
  {author} {\bibfnamefont {D.~E.}\ \bibnamefont {Krause}}, \ and\ \bibinfo
  {author} {\bibfnamefont {V.~M.}\ \bibnamefont {Mostepanenko}},\ }\Doi
  {10.1103/PhysRevD.75.077101} {\bibfield  {journal} {\bibinfo  {journal}
  {Phys. Rev. D},\ }\textbf {\bibinfo {volume} {75}},\ \bibinfo {eid} {077101}
  (\bibinfo {year} {2007})}\BibitemShut {NoStop}%
\end{thebibliography}%

\end{document}